\UseRawInputEncoding

\documentclass{nature}

\usepackage{comment}

\usepackage{amsmath}
\usepackage[dvipsnames]{xcolor}
\usepackage{soul}
\usepackage{graphicx}
\usepackage{amssymb}
\usepackage{pifont}
\usepackage{longtable}
\usepackage{multirow}
\usepackage{arydshln}
\usepackage{hyperref}
\usepackage{caption}

\usepackage[commandnameprefix=always]{changes}

\usepackage[super,comma,sort&compress]{natbib}
\usepackage{url}

\usepackage[switch, modulo]{lineno}
\modulolinenumbers[1]

\title{Deep learning-driven likelihood-free parameter inference for 21-cm forest observations}

\author{Tian-Yang Sun$^{1}$,
Yue Shao$^{1}$,
Yichao Li$^{1}$,
Yidong Xu$^{2,3}$,
He Wang$^{4,5}$
\& Xin Zhang$^{1,6,7,\star}$
}

\begin{document}

\maketitle

\begin{affiliations}
 \item Liaoning Key Laboratory of Cosmology and Astrophysics, College of Sciences, Northeastern University, Shenyang 110819, China
 \item National Astronomical Observatories, Chinese Academy of Sciences, Beijing 100101, China
 \item Key Laboratory of Radio Astronomy and Technology, Chinese Academy of Sciences, Beijing 100101, China
 \item International Centre for Theoretical Physics Asia-Pacific, University of Chinese Academy of Sciences, Beijing 100049, China
 \item Taiji Laboratory for Gravitational Wave Universe (Beijing/Hangzhou), University of Chinese Academy of Sciences, Beijing 100049, China
 \item National Frontiers Science Center for Industrial Intelligence and Systems Optimization, Northeastern University, Shenyang 110819, China
 \item MOE Key Laboratory of Data Analytics and Optimization for Smart Industry, Northeast-ern University, Shenyang 110819, China

\end{affiliations}

\begin{abstract}
The hyperfine structure absorption lines of neutral hydrogen in high-redshift radio spectra, known as the 21-cm forest, have been demonstrated through simulations as a powerful probe of small-scale structures governed by dark matter (DM) properties and the thermal history of intergalactic medium (IGM). By measuring the one-dimensional power spectrum of the 21-cm forest, parameter degeneracies can be broken, offering key constraints on the properties of both DM and IGM. However, conventional methods are hindered by computationally expensive simulations and a non-Gaussian likelihood function. To overcome these challenges, we propose a deep learning approach that combines generative normalizing flows for data augmentation and inference normalizing flows for parameter estimation, enabling accurate results from minimally simulated datasets. Using mock data from the Square Kilometre Array, we demonstrate the capability of  this deep learning-driven approach to generating posterior distributions, providing a robust tool for probing DM and the cosmic heating history.
\end{abstract}

\newpage

\section*{Introduction}
The hyperfine structure transition of neutral hydrogen atoms from the epoch of reionization (EoR) creates narrow absorption lines on spectra of high-redshift radio bright sources (e.g. radio-loud quasars), with a rest-frame wavelength of 21 cm. The 21-cm absorption lines from various structures at different distances along the line of sight are redshifted to different frequencies, giving rise to forest-like features on the spectra, which is called the 21-cm forest~\citep{Carilli2002,Furlanetto:2002ng,Furlanetto:2006dt,Xu2009,Xu:2010us,Ciardi2013} in analogy to the Lyman-$\alpha$ (Ly-$\alpha$) forest at lower redshifts. Simulations suggest that the 21-cm forest could serve as a sensitive probe of the cosmic heating history~\citep{Xu2009,Xu:2010us}. After the formation of the first galaxies, the generated light not only ionizes hydrogen atoms but also heats the gas simultaneously (with X-rays being the primary heating source). As the spin temperature of the gas increases, the absorption at 21 cm weakens (reducing the optical depth), thus causing the 21-cm forest signal to become weaker. In principle, the strength of the signal can be used to measure the cosmic heating history, thereby gaining insight into the properties of the first galaxies.

On the other hand, understanding the properties of dark matter (DM) hinges crucially on probing small-scale structures of the universe. The cold DM (CDM) leads to an excess of structure on small scales, while warm DM (WDM) or ultra-light DM results in suppressed structure formation on small scales~\citep{Bode:2000gq,Hu:2000ke}.
Being sensitive to the small-scale structures, it has been shown that the 21-cm forest can be used to probe the minihalos and dwarf galaxies during the EoR~\citep{Xu2010,Xu:2010us}, so as to provide precise measurements of small-scale cosmic structures, and can be utilized to measure the mass of DM particles~\citep{Shimabukuro:2014ava}.

During the Square Kilometre Array (SKA) era, conducting 21-cm forest observations theoretically allows the mass of DM particles to be constrained based on the number density of absorption lines~\citep{Furlanetto:2002ng,Shimabukuro:2014ava}. However, detecting the 21-cm signal is by no means an easy task. In particular, if the extent of cosmic heating is too strong, it will render the 21-cm forest signal undetectable, thereby making it impossible to measure either the properties of DM particles or the temperature of the intergalactic medium (IGM) through this method.

Our recent work has proposed an effective solution to this problem, namely, measuring the one-dimensional (1D) power spectrum of the 21-cm forest~\citep{Shao:2023agv}. This approach not only allows for the extraction of faint signals but also breaks the degeneracy between the mass of DM particles and the temperature of the IGM, enabling both to be measured simultaneously. This is because, when transforming from frequency space to $k$-space, the scale-dependence characteristics of the 21-cm forest signal can be displayed, while noise can be effectively suppressed as it lacks such scale dependence, enabling even faint signals to be extracted. Furthermore, the properties of DM particles (whether warm or cold) and the temperature of the IGM mainly affect the shape and amplitude of the power spectrum, respectively. Thus, both can be simultaneously constrained through power spectrum measurements. If this approach can be successfully implemented during the SKA era, it will be significant for research of DM and first galaxies.

To achieve this goal, a series of challenges still need to be overcome. Observationally, it requires a sufficiently sensitive experiment and long observation times to detect the weak signals, as well as a sufficient number of high-redshift radio-bright point sources to suppress the cosmic variance~\citep{Niu:2024eyf}. Additionally, even if we can achieve 21-cm forest observations, it is not straightforward to constrain the properties of DM particles and the cosmic heating history, because currently there is no analytical or empirical model to connect the parameters (such as the mass of DM particles and the temperature of the IGM) with observables (the 1D power spectrum of the 21-cm forest). In this case, to achieve parameter inference, we can only rely on a large number of simulations spanning a sufficient large parameter space to calculate observables, and then  construct a likelihood function, thus utilizing Bayesian methods for parameter inference~\citep{Bird:2023evb}. This is how parameter inference is done for constraining DM particles using the Ly-$\alpha$ forest~\citep{Baur:2015jsy}. Such a large number of simulations require a significant amount of computing resources, making it both very expensive and extremely time-consuming. Moreover, the computational costs for the 21-cm forest simulations are considerably higher due to its major contributions from smaller scales ($k$ as high as $\sim 100$ Mpc$^{-1}$), compared to the Ly-$\alpha$ forest~\citep{Carilli:2002ky,Furlanetto:2002ng}.

Additional challenges arise from the non-Gaussian likelihood function of the 21-cm forest signals induced by the nonlinear effect of gravitational clustering~\citep{Mondal:2016hmf} and the non-local effect of radiation~\citep{Munoz:2023kkg}. Due to the non-Gaussian likelihood function for the 1D power spectrum, traditional Bayesian methods for parameter inference are somewhat inadequate, potentially leading to severely distorted inference results. Although methods like traditional Gaussian mixture models can fit likelihood functions under non-Gaussian conditions~\citep{press1992vetterling,mclachlan2000wiley}, they require significant computational resources, making them less practical for extensive data analysis. Additionally, the correlation between the 1D power spectrum at different scales also affects the precision of parameter inference framework, as the coupling of errors across scales increases the complexity of modeling the likelihood.

This work proposes a set of solutions to these issues, allowing  parameter inference of 21-cm forest observations without the need for an explicit model or the construction of the likelihood function, using only a limited number of simulated samples. This enables us to obtain constraints on the properties of DM particles and the history of cosmic heating efficiently from 1D power spectrum measurements of the 21-cm forest. The key strategy relies on variational inference methods based on deep learning~\citep{blei2017variational}, especially the technique of normalizing flows (NFs)~\citep{rezende2015variational}. These methods have found various applications in fields such as the large-scale generation of neutral hydrogen sky maps~\citep{Hassan:2021ymv,Friedman:2022lds}, parameter inference during the EoR using 21-cm tomography~\citep{Gillet:2018fgb,Zhao:2021ddh,Zhao:2022ren,Zhao:2023tep}, inference of gravitational wave source parameters under the influence of noise transients~\citep{Sun:2023vlq,Xiong:2024gpx}, and the inference of the IGM thermal parameters based on the Ly-$\alpha$ forest~\citep{Nayak:2023tyh,Maitra:2024ggb}. Within this framework, not only can a large number of simulated samples be easily generated through generative neural networks, but parameter inference can also be achieved without a likelihood function. This approach addresses the aforementioned challenges of large sample sizes and a non-Gaussian likelihood function, respectively, for 21-cm forest observations, paving the way for advancing the usage of this probe to constrain the fundamental physics during the cosmic dawn.

\section*{Results}
\subsection{Simulation setup.}

\begin{figure}
\centering
\includegraphics[angle=0, width=16.0cm]{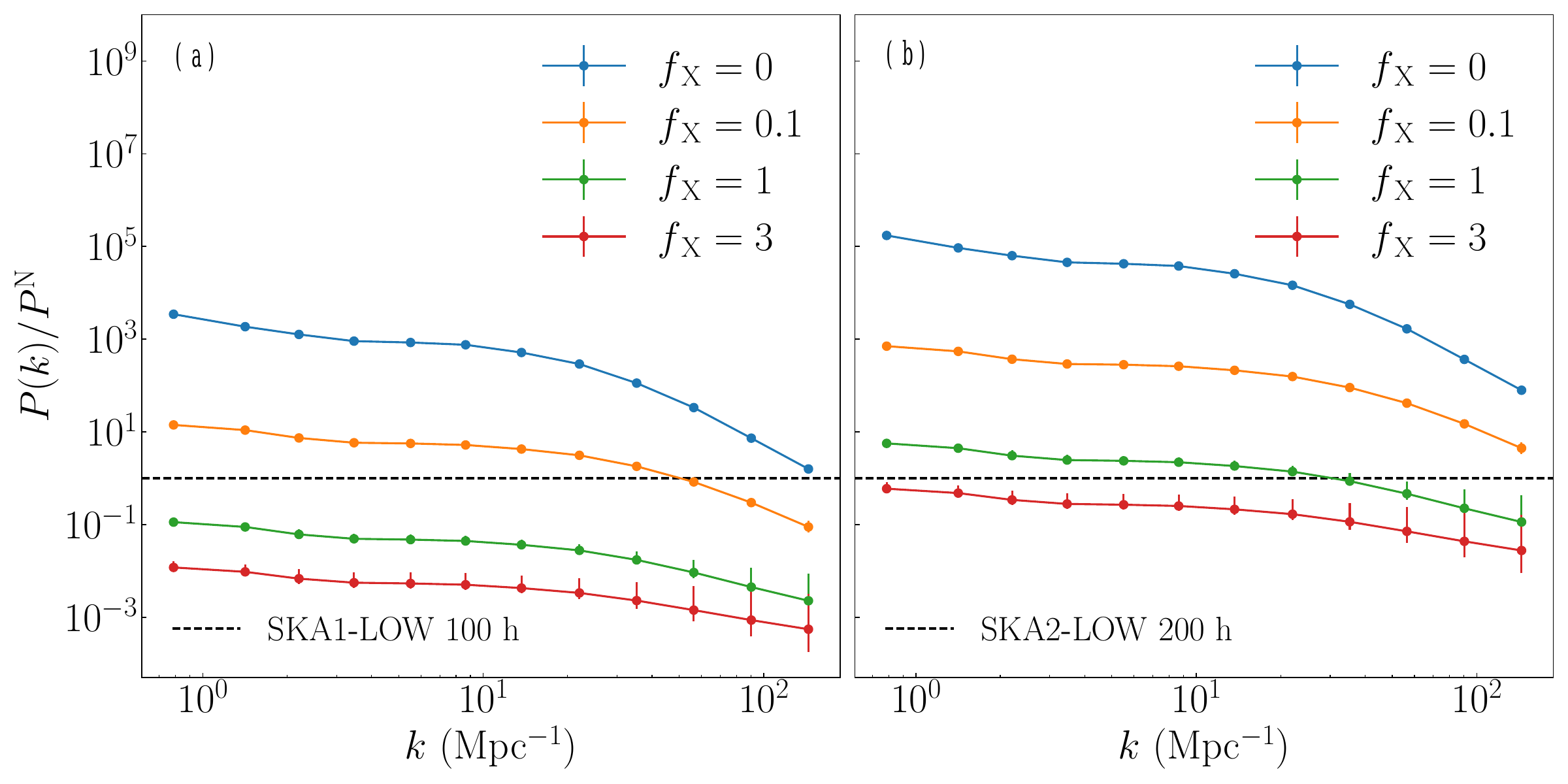}
\caption{\label{fig SNR}
  \textbf{The SNR of the 1D power spectrum of the 21-cm forest corresponding to different heating levels in the WDM model with $m_{\rm WDM} = 6~{\rm keV}$.}
 The blue, orange, green and red curves correspond to $f_{\rm X} = 0$, $f_{\rm X} = 0.1$, $f_{\rm X} = 1$ and $f_{\rm X} = 3$, respectively. \textbf{a} Depicts the SNR with an integration time of $100$ h on each source using SKA1-LOW. \textbf{b} Depicts the results for an integration time of $200$ h on each source using SKA2-LOW. The error bars show the sample variances of the 1D power spectra of the 21-cm forest. The black dashed line in each panel indicates the SNR of 1.
  }
\end{figure}

We take the same approach as Ref.~\citep{Shao:2023agv}, and simulate the 21-cm forest signals from the EoR for various X-ray production efficiency of the first galaxies parameterized by $f_{\rm X}$, and for various DM particle masses of $m_{\rm WDM}$ (see Methods). The parameter $f_{\rm X}$ controls the efficiency of X-ray heating in the early universe, directly influencing the IGM temperature $T_{\rm K}$. Given the current Hydrogen Epoch of Reionization Array (HERA) findings, $T_{\rm K}$ is constrained to 15.6 K $< T_{\rm K} <$ 656.7 K at $z\sim 8$ with 95\% confidence~\citep{HERA:2022wmy}, which demarcates the rough signal level of the 21-cm forest constrained by the heating effect. Therefore, our analysis mainly focuses on two characteristic heating levels for discussion: the case with a weaker heating effect ($T_{\rm K} \approx$ 60 K at redshift 9 corresponding to $f_{\rm X} \sim 0.1$) and the case with a stronger heating effect ($T_{\rm K} \approx$ $600$ K at redshift 9 corresponding to $f_{\rm X} \sim 1$). As $f_{\rm X}$ increases, $T_{\rm K}$ rises, regulating the X-ray heating in the early universe and strongly influencing the detectability of the 21-cm forest signal. For comparison, we also simulate cases with $f_{\rm X} = 0$ representing an unheated IGM and $f_{\rm X} = 3$ for extremely high productivity of X-rays in the early universe. For the DM properties, although some recent astrophysical observations suggest a lower limit on the WDM particle mass ($m_{\rm WDM}$) of approximately 6 kilo electronvolts (keV), models with lower WDM particle mass have not been definitively excluded~\citep{Garzilli:2019qki, Enzi:2020ieg, Villasenor:2022aiy, Zelko:2022tgf, Irsic:2023equ}. Furthermore, as the value of $m_{\rm WDM}$ increases, the structure formation characteristics of WDM models on small scales become increasingly similar to those of CDM model. We therefore simulate WDM models with mass in the several keV range for subsequent analysis.
Assuming a fiducial WDM model with $m_{\rm WDM}=6$ keV, Figure~\ref{fig SNR} shows the signal-to-noise ratio (SNR) of the 1D power spectrum of the simulated 21-cm forest, represented by $P(k)/P^{\rm N}$, where $P(k)$ is the 1D power spectrum of the pure 21-cm forest signal and $P^{\rm N}$ is thermal noise of the 1D power spectrum, with different heating levels for the noise levels corresponding to phase-one and phase-two low-frequency arrays of the SKA (SKA1-LOW~\cite{Acedo:2020lve} and SKA2-LOW~\cite{Braun:2019gdo,SKAO}), respectively. Our results indicate that a lower heating level ($T_{\rm K} = 60~\rm K$ at $z = 9$) allows for measurement of the 1D power spectrum of the 21-cm forest with an integration time of $100$ h per background source with a flux density of $10$ mJy using the SKA1-LOW. In contrast, a higher heating level ($T_{\rm K} = 600~ \rm K$ at $z = 9$) necessitates 200 h on each source with the same flux density using SKA2-LOW.

Given the high dynamic range required for the 21-cm forest simulations, the simulation-based Bayesian methods for parameter inference are inefficient and almost infeasible, considering the requirement of computational time and resources. For instance, simulating the density field for $m_{\rm WDM} = 6$ keV requires approximately $600$~h on $64$ cores (Intel(R) Xeon(R) Gold 6271C CPU @ 2.60GHz). To achieve parameter inference with minimal computational expenditure, we first simulate the 21-cm forest signals for a limited sampling of the parameter space. Specifically, $m_{\rm WDM}$ ranges from $3$ keV to $9$ keV with intervals of $1$ keV, while $T_{\rm K}$ ranges from $40$ K to $80$ K with intervals of $5$ K for weaker heating levels, and from $400$ K to $800$ K with intervals of $100$ K for stronger heating levels. This parameter space results in a dataset of $567000$ samples (corresponding to $7$ values of $m_{\rm WDM}$ and $9$ values of $T_{\rm K}$, for a total of $7 \times 9 \times 9000 = 567000$ samples; see Methods) for weaker heating levels, which, while sufficiently large, is unevenly distributed and does not fully cover the entire parameter space. To ensure robustness, the validation dataset, split from the training data, is employed to fine-tune model parameters and prevent overfitting. The testing dataset, also separated from the initial dataset, is reserved exclusively for evaluating the model's predictive performance on unseen data.

\begin{figure}
\centering
\includegraphics[angle=0, width=16.0cm]{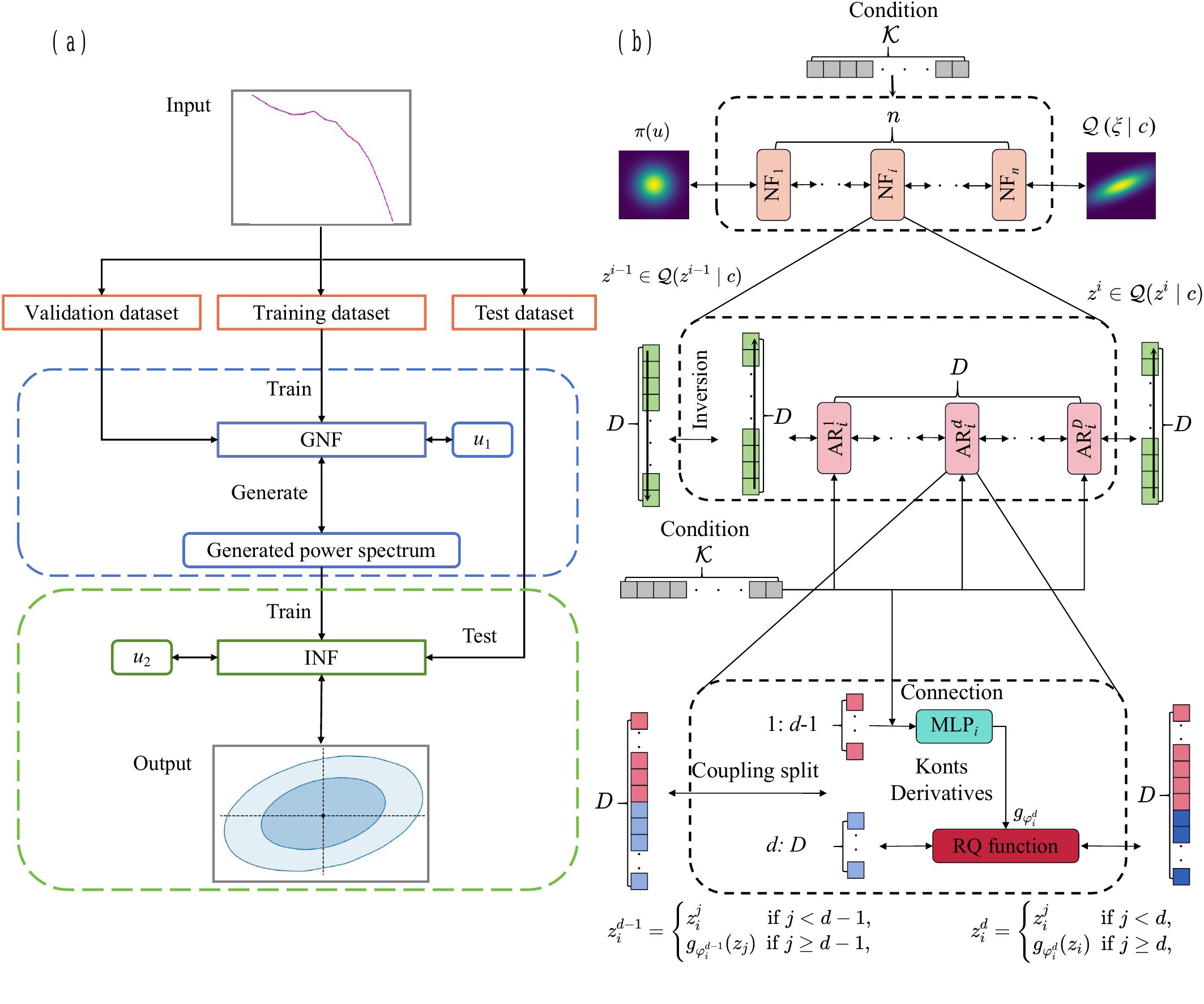}
\caption{\label{fig NSF}
  \textbf{Network workflow and schematic diagram of each part structure}. \textbf{a} Depicts network workflow chart. \textbf{b} Depicts schematic diagram of the architecture of the NF.  Both the GNF and INF are composed of RQ-NSF (AR). In the GNF, the conditions are parameters, and the output is a 1D power spectrum. Conversely, in the INF, the condition is a 1D power spectrum, and the outputs are parameters. The network consists of a sequence of flows, each of which consists of a sequence of autoregressive layers that share parameters. Input the condition and the sampling vector obtained from the base distribution into the NFs to generate the output. The simulated power spectra data set is divided into a training set, a validation set, and a testing set. The training set and validation set are used to train the GNF, which generates reasonable power spectra for any given parameters. The INF is trained using the 1D power spectra generated by the GNF to learn the parameter distribution corresponding to any given power spectrum. Different testing sets of simulated power spectra are used to verify the rationality of the GNF and the INF, respectively.
  }
\end{figure}

As shown in Supplementary Fig.~\ref{sfig1}, this non-uniform sampling of the parameter space impairs likelihood-free inference methods, leading to inaccurate probability estimations in under-sampled regions of the parameter space. However, as presented in Supplementary Fig.~\ref{sfig2}, while the parameter space distribution is irregular, the corresponding power spectrum space exhibits continuity, indicating that the power spectra are less sensitive to gaps in parameter sampling. To leverage this property, our framework combines a generative normalizing flow (GNF) to fill the gaps in the parameter space and an inference normalizing flow (INF) to accurately infer parameters, as shown in Fig.~\ref{fig NSF}.

\subsection{Data augmentation.}

\begin{figure}
\centering
\includegraphics[angle=0, width=16.0cm]{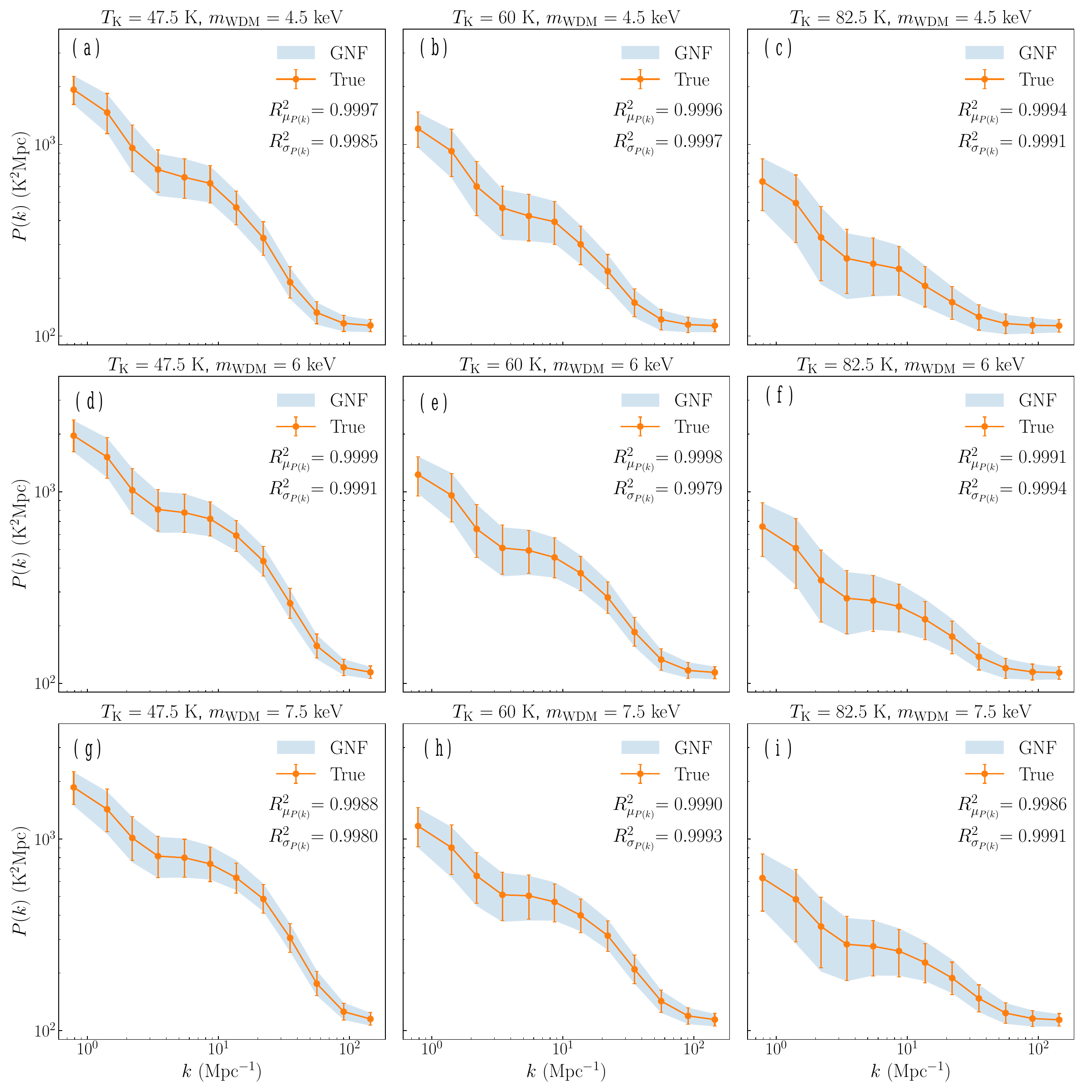}
\caption{\label{fig PS generation}
  \textbf{The mean values and $1\sigma$ distribution of the GNF-generated 1D power spectra and simulated 1D power spectra.}
  The data are derived from results of different $T_{\rm K}$ and $m_{\rm WDM}$, with an integration time of $100$ h using SKA1-LOW.
   \textbf{a - c}, \textbf{d - f}, \textbf{g - i} Present the 1D power spectra of 21-cm forest with $m_{\rm WDM} = 3$ keV, $m_{\rm WDM} = 6$ keV and $m_{\rm WDM} = 9$ keV, respectively. The 1D power spectra of the 21-cm forest are presented in \textbf{(a, d, g)}, \textbf{(b, e, h)} and \textbf{(c, f, i)} with $T_{\rm K} = 47.5$ K, $T_{\rm K} = 60$ K and $T_{\rm K} = 82.5$ K, respectively. The blue shaded area is the $1\sigma$ range of the 1D power spectrum generated by the GNF. The orange line and bar are the mean and $1\sigma$ error bar of the simulated power spectrum. $R_{\mu_{P(k)}}^{2}$ and $R_{\sigma_{P(k)}}^{2}$ are the correlation coefficients of the mean and standard deviation, respectively.
  }
\end{figure}

To guarantee a robust parameter inference, we first assess the reliability of the 1D power spectrum of the 21-cm forest generated by the GNF. Figure~\ref{fig PS generation} compares the mean value and standard deviation of the 1D power spectra generated by the GNF with those from simulations. The results confirm that the GNF consistently reproduces the 1D power spectra generated from the training set as presented in \textbf{(b, e, h)} of Fig.~\ref{fig PS generation}. In addition, it effectively generates the 1D power spectra for both interpolated and extrapolated parameter regions as presented in \textbf{(a, d, g)} and \textbf{(c, f, i)} of Fig.~\ref{fig PS generation}, demonstrating its generalization capability. To quantify the reasonableness of the GNF model, we applied the coefficient of determination ($R^{2}$) to measure the correlation between the 1D power spectra generated by the GNF and those from simulations. Achieving $R^{2} > 0.99$ indicates a high reliability of the GNF in reproducing these 1D power spectra. This examination, combining statistical measures and visual inspections, confirms the reliability of the GNF model in generating the 1D power spectrum of the 21-cm forest. Additionally, using the GNF, approximately $24000$ 1D power spectra can be generated in about $1$ s, a speed unattainable by traditional simulations. In contrast, generating the same number of samples with traditional simulations would require approximately $280$ h.

\begin{figure}
\centering
\includegraphics[angle=0, width=16.0cm]{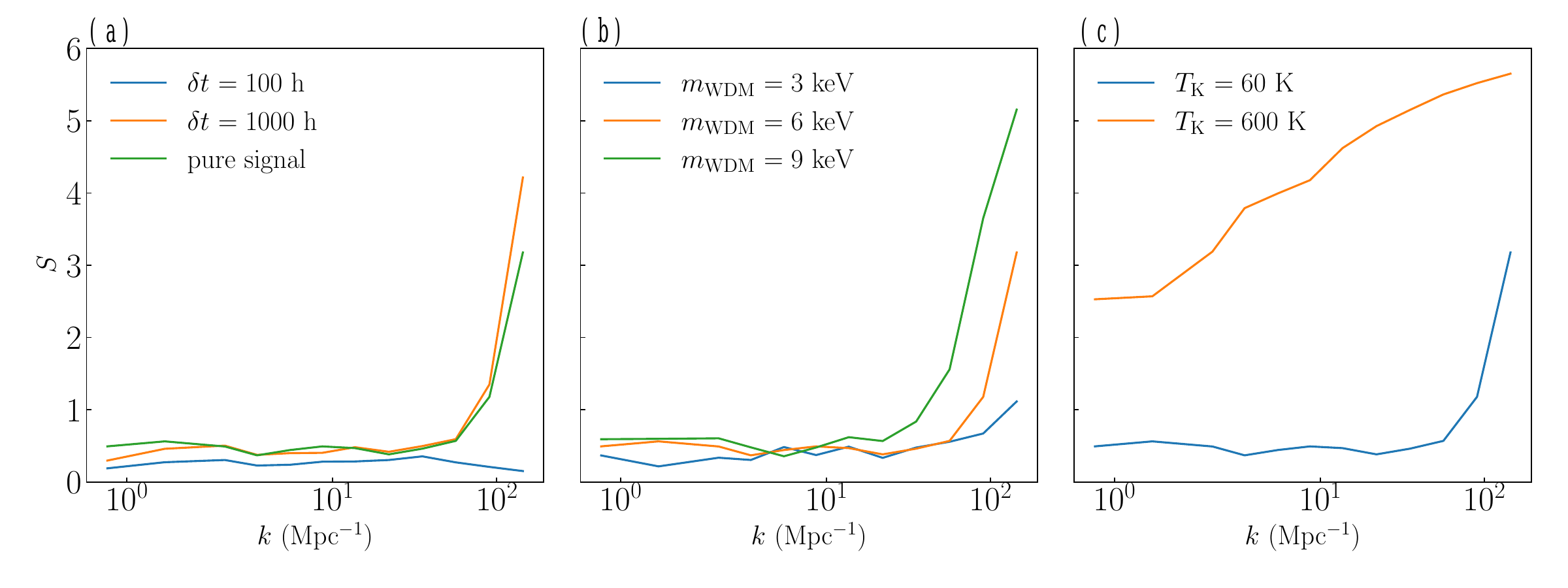}
\caption{\label{fig data distribution S}
  \textbf{The skewness coefficient ($S$) of the 1D power spectrum at different integration time, $m_{\rm WDM}$, and $T_{\rm K}$.}
The blue, orange, and green curves in \textbf{(a)} present the $S$ of the 1D power spectrum at different $k$ when $m_{\rm WDM} = 6$ keV and $T_{\rm K} =$ $60$ K, and the SKA1-LOW integration time is $100$ h, $1000$ h, and infinity (pure signal), respectively. The blue, orange, and green curves in \textbf{(b)} present the $S$ of the 1D power spectra at different $k$ when $T_{\rm K} =$ $60$ K, and $m_{\rm WDM}$ is $3$ keV, $6$ keV, and $9$ keV, respectively. The blue and orange curves in \textbf{(c)} present the 1D power spectrum $S$ of the pure signal at different $k$ when $m_{\rm WDM} = 6$ keV and $T_{\rm K}$ is $60$ K and $600$ K, respectively.
  }
\end{figure}

We used the skewness coefficient ($S$) to describe the asymmetry of the 1D power spectrum at each $k$ in Fig.~\ref{fig data distribution S}. It shows that at smaller scales and with longer integration times, deviations from a Gaussian distribution increase. This trend is attributed to the increasing dominance of small-scale strong nonlinear effects compared to thermal noise. Additionally, we observed gradual deviations from a normal distribution with increasing $m_{\rm WDM}$ and $T_{\rm K}$. These deviations are attributed to the increased number of halos with higher $m_{\rm WDM}$, which significantly alters the density field, and to temperature variations influencing the 1D power spectrum of the 21-cm forest as $T_{\rm K}$ increases. Additionally, we quantified the similarity between the distributions generated by the GNF and the simulation in Supplementary Fig.~\ref{sfig3} using the Jensen-Shannon divergence (JSD)~\citep{61115,rao1987differential}, a symmetric measure of similarity between two probability distributions. Our analysis revealed that the JSD value is less than $0.05$, indicating a high degree of similarity between the distributions generated by the GNF and the simulation. As previously noted with JSD and $S$ analyses, these results confirm that under different heating levels, the 1D power spectrum of the 21-cm forest observed with an integration time of $100$~h using SKA1-LOW and those measured with an integration time of $200$~h using SKA2-LOW exhibit Gaussian and non-Gaussian distributions, respectively. This highlights the capability of the GNF in modeling the 1D power spectrum of the 21-cm forest under various astrophysical conditions.

The correlation coefficient $\rho_{ij}$ between different $k$ bins of the 1D power spectrum of the 21-cm forest is also crucial for accurately estimating the uncertainties in parameter inference. Here, $i$ and $j$ refer to different $k$ bins. Ignoring the correlation in the 1D power spectrum between these different $k$ bins can lead to incorrect error estimates. In Supplementary Fig.~\ref{sfig4}, we show the values of $\rho_{ij}$ of the pure signal under different conditions. We find that the correlations at small scales are significant, especially in scenarios with strong heating effects. However, considering the impact of SKA thermal noise in actual observations, as shown in Supplementary Fig.~\ref{sfig5}, the dominance of thermal noise at small scales reduces the overall correlation strength. Nonetheless, the correlations remain non-negligible. Therefore, although combining GNF with Bayesian inference allows us to obtain posterior distributions, the aforementioned correlations significantly increase the complexity of modeling the likelihood function. As a result, generating $2^{16}$ posterior samples requires several days of computation on $32$ cores.

\subsection{Parameter inference.}

\begin{figure}
\centering
\includegraphics[angle=0, width=16.0cm]{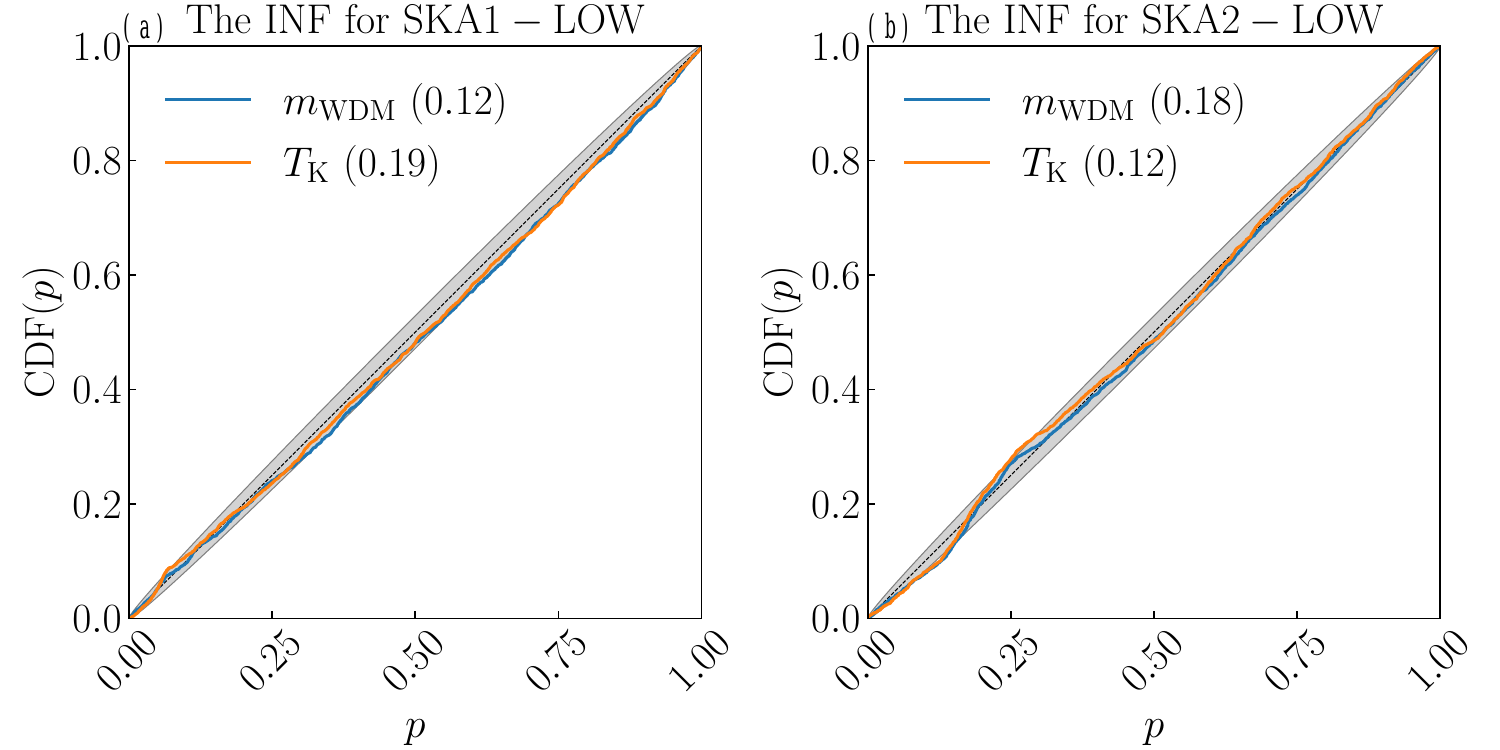}
\caption{\label{fig p_p}
  \textbf{P-P plot of the INF parameter inference.}
     \textbf{a} Depicts the P-P plot of the INF inference with an integration time of $100$ h using SKA1-LOW and for a total of $2700$ power spectra with $m_{\rm WDM}$ ranging from $5$ keV to $7$ keV and $T_{\rm K}$ from $50$ K to $70$ K. \textbf{b} Depicts the P-P plot of the INF inference with an integration time of $200$ h using SKA2-LOW and for a total of $2700$ power spectra with $m_{\rm WDM}$ ranging from $5$ keV to $7$ keV and $T_{\rm K}$ from $400$ K to $800$ K. The gray shaded area is the range of the $3\sigma$ confidence interval. KS test $p$-values are denoted in the legend.
  }
\end{figure}

To assess the performance of the INF in deriving cosmological parameters, we evaluate two key aspects: the reliability of the inferred parameters and the precision of the parameter estimates. First, we need to confirm the reliability of the INF inference parameters. We conducted the Kolmogorov-Smirnov (KS) test~\citep{Lopes2011}. The KS test is a non-parametric test that compares the cumulative distribution function (CDF) of the sample data to a reference distribution. By computing the CDF of the percentiles of the true values of each parameter derived from the marginal 1D posterior distribution, we can evaluate the model's performance. For each 21-cm forest 1D power spectrum and the corresponding confidence interval of the posterior distribution, the percentile values of the injected parameters are computed. Ideally, the CDF for each parameter should be close to the diagonal, as the percentiles should be evenly distributed between 0 and 1. We measured the CDF of the true parameters falling within specific confidence levels as presented in Fig.~\ref{fig p_p}. The results demonstrated the CDF curves within the $3\sigma$ confidence bands, with $p$-values greater than $0.05$ for each parameter. A KS test $p$-value greater than $0.05$ indicates that there is no significant difference between the CDF of the sample data and the reference distribution, confirming that the observed deviations are within the expected range due to random sampling variability. These findings affirm the validity and accuracy of our inference method within the expected error ranges, providing confidence in its application for cosmological research.

Next, we determine the precision of the parameters inferred by INF. Similar to findings in Refs.~\citep{Nayak:2023tyh, Zhao:2021ddh},
we encounter challenges due to limited prior parameter bounds, as probabilities outside the predefined range cannot be estimated. However, this can be solved by expanding the coverage of the simulated parameter space. 
Additionally, using reciprocal parameter values in transformations could improve the inference framework's applicability, making it possible to avoid simulating to infinity. Within the central regions of the parameter space, our analysis reveals distinct patterns in the relative error $\varepsilon$ of parameter inference for $m_{\rm WDM}$ and $T_{\rm K}$ under various heating levels, as presented in Supplementary Fig.~\ref{sfig6}. Here, the relative error, defined as the precision, is given by $\varepsilon(\xi ) = \sigma (\xi ) / \xi$, where $\xi$ represents the true parameter ($m_{\rm WDM}$ or $T_{\rm K}$). Although SKA2-LOW is more sensitive, the higher error is due to the fact that our discussion for SKA1-LOW focused on cases with weaker heating effects, while increasing the heating effect suppresses the 21-cm forest signal, leading to larger inference errors. We find that $T_{\rm K}$ significantly affects the precision of $m_{\rm WDM}$ estimates. Conversely, $m_{\rm WDM}$ has a relatively small impact on the precision of $T_{\rm K}$ estimates. This is because $T_{\rm K}$ influences the 1D power spectrum values across all scales, while $m_{\rm WDM}$ primarily affects the 1D power spectrum values on small scales, as shown in Supplementary Fig.~\ref{sfig2}. Therefore, as presented in Supplementary Fig.~\ref{sfig7}, $T_{\rm K}$ can achieve high precision by relying solely on large-scale information, almost unaffected by $m_{\rm WDM}$. On the other hand, $m_{\rm WDM}$ mainly relies on small-scale and medium-scale information, which is influenced by $T_{\rm K}$. Therefore, the 1D power spectrum of different scales will break the degeneracy, resulting in a significant improvement in the precision of parameter estimates. This dependence highlights the interplay between these parameters and emphasizes the ability of our network to effectively constrain each parameter. Generating $2^{16}$ posterior samples with INF takes only $32.2$ ms, showcasing its remarkable efficiency, far surpassing the speed of GNF combined with Bayesian inference.

\begin{figure}
\centering
\includegraphics[angle=0, width=16.0cm]{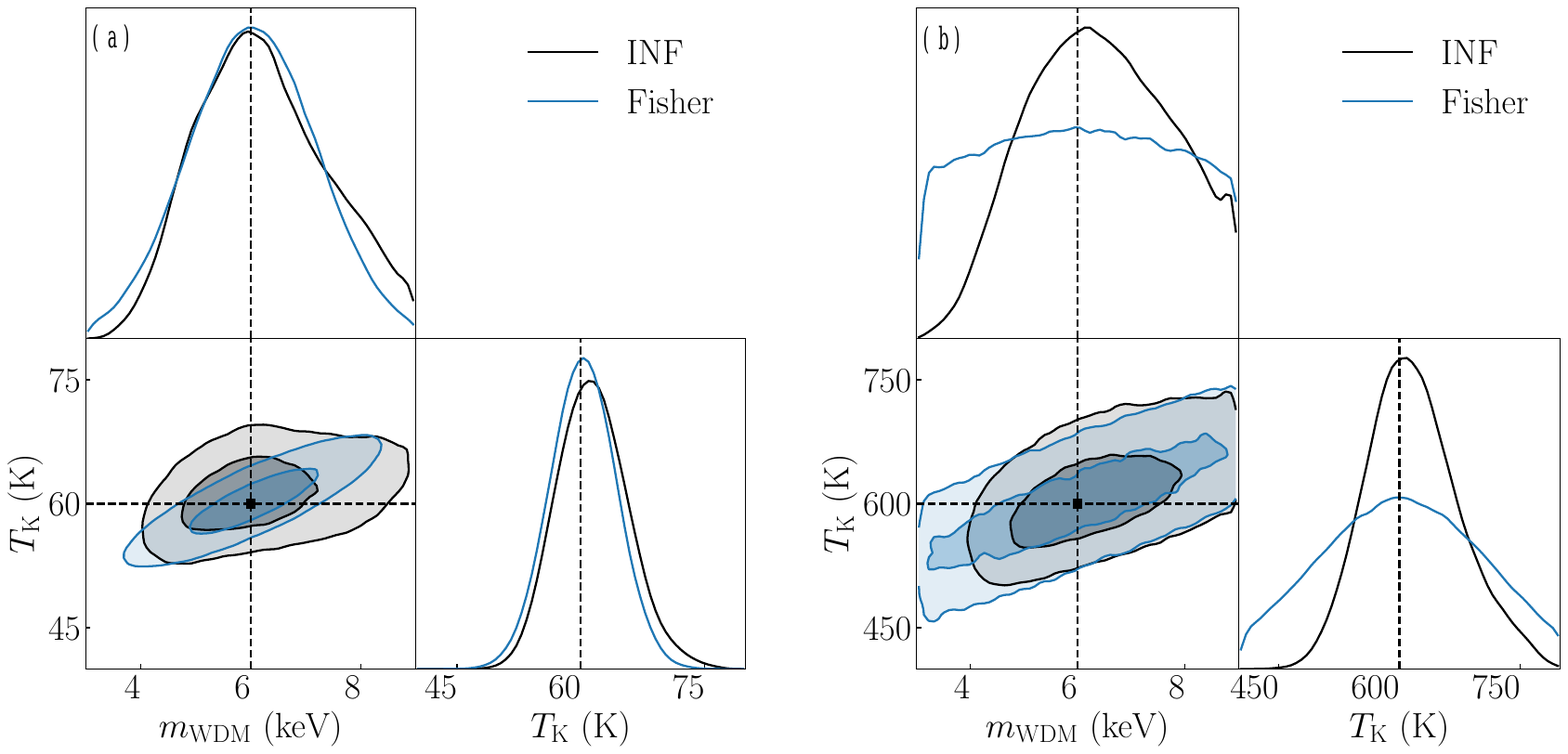}
\caption{\label{fig conter}
\textbf{The posterior distributions for $m_{\rm WDM}$ and $T_{\rm K}$ with the INF and Fisher matrix.}
The black contour is based on the INF and the blue contour is based on the Fisher matrix. \textbf{a} Depicts the results when $m_{\rm WDM} = 6$ keV and $T_{\rm K} =$ $60$ K with an integration time of $100$ h using SKA1-LOW. \textbf{b} Depicts the results when $m_{\rm WDM} = 6$ keV and $T_{\rm K} = 600$ K with an integration time of $200$ h using SKA2-LOW. Contours represent $1\sigma$ and $2\sigma$ confidence intervals.}
\end{figure}

For low heating level conditions, we utilized SKA1-LOW with an integration time of $100$ h to examine the constraint results of posterior parameter inference, assuming $m_{\rm WDM} = 6$ keV and $T_{\rm K} = 60$ K. We compared the precision of the posterior inference results from the INF with the range predicted by the Fisher matrix as presented in Fig.~\ref{fig conter}. The INF method successfully recovered the posterior distribution of parameters, with the true values lying within the estimated $1\sigma$ confidence region. This demonstrates precision comparable to the predictions of the Fisher matrix. The final average constraint result of the 1D power spectra of the 21-cm forest for all simulations are $m_{\rm WDM} = 6.28_{-1.59}^{+1.61}$ keV and $T_{\rm K} = 60.17_{-6.66}^{+6.16}$ K. This confirms the reliability of the INF method in achieving precise parameter inference under low heating conditions.

Under high heating level conditions, using SKA2-LOW with an integration time of $200$ h ($m_{\rm WDM}=6$ keV, $T_{\rm K}=600$ K), we observed significantly different results. Although both the Fisher matrix and INF methods presented consistent degeneracy directions, the Fisher matrix approach failed to adequately constrain the parameters due to the non-Gaussian likelihood function of the 1D power spectrum of the 21-cm forest. This highlights the effectiveness of likelihood-free inference methods like the INF in such scenarios. The final average constraint results of the 1D power spectra for all simulations are $m_{\rm WDM} = 6.47_{-1.88}^{+2.04}$ keV and $T_{\rm K} = 627.95_{-91.07}^{+94.18}$ K. These results demonstrate that, while the Fisher matrix approach is inadequate under a non-Gaussian likelihood function, the INF method remains effective. 

\section*{Discussion}
Observing the 21-cm forest signals during the EoR provides a valuable opportunity to gain insight into DM properties and the heating history in the first billion years of the universe. However, due to the lack of explicit relation between the physical parameters and the observables, parameter inference has been impeded by the limited parameter space covered by computationally expensive simulations and the infeasible requirement on computational resources. Moreover, the non-Gaussian likelihood function of the 21-cm forest signal complicates the evaluation of the likelihood function, resulting in biased inference of parameters. In this work, we introduce a deep learning-based approach that does not rely on traditional analytical models, enabling inference of $T_{\rm{K}}$ and $m_{\rm WDM}$ with a small number of simulations, unlike analytical models that establish an explicit relationship between $m_{\rm WDM}$, $T_{\rm{K}}$ during the reionization period, and the statistical characteristics of the 21 cm signal. Furthermore, this approach could extend to other non-Gaussian scenarios, allowing for parameter inference using arbitrary 21-cm forest signals.

For data analysis, we calculate the two-point correlation function of the 21-cm forest and convert it to the $k$-space, resulting in the 1D power spectrum along the line of sight. The choice stems from the faintness of the 21-cm forest signal, making a power spectrum analysis more beneficial. Nonetheless, other studies~\citep{Zhao:2023tep} suggest alternative methods that could more effectively compress data and yield more precise parameter constraints. Future research will explore these methods for nearly lossless parameter inference, enhancing the effectiveness of constraints. Additionally, we plan to incorporate factors like metal absorption lines~\citep{Bhagwat:2022uda} coupling and foreground to create a more accurate dataset. To  improve the astrophysical modeling, the inclusion of spatial temperature non-uniformity and X-ray source distribution variations~\citep{Pritchard:2006sq,Park:2018ljd} is also worth considering. Additionally, the impact of including additional parameters that influence the 21-cm forest warrants  investigation, where methods like Latin hypercube sampling (LHS) could potentially enhance sampling efficiency~\citep{mckay2000comparison,Schmit:2017pho}.

Overall, our research demonstrates that deep learning-based methods, particularly NFs, offer a promising approach to analyzing the 21-cm forest signal. These methods address the challenges of simulation data scarcity and complexity, providing robust and accurate parameter inference. In addition, the INF method can assist the 21-cm forest in distinguishing between the CDM and WDM models. Given the computationally intensive nature of CDM simulations, we conservatively use the simulated upper limit of $9$ keV as a stand-in for CDM in our analysis. In Supplementary Fig.~\ref{sfig8}, we assess the discriminative power of the INF method by comparing the confidence intervals with the upper limit ($9$ keV) of the parameter space for the simulated $m_{\rm WDM}$ under different $m_{\rm WDM}$ and $T_{\rm K}$. The results indicate that the INF method can distinguish the 1D power spectrum of $m_{\rm WDM}$ values less than 6 keV from that of the CDM model at a $2\sigma$ confidence level. more, the deep learning-based parameter inference method demonstrated here not only advances astronomy and cosmology but also shows potential for handling complex and irregular datasets~\citep{more2016survey}. In fields like biomedical research, where sample scarcity hinders the study of rare diseases~\citep{banerjee2023machine}, our method can significantly improve data analysis capabilities. As the SKA era approaches, these innovative techniques will be invaluable for extracting maximum scientific information from 21-cm observations, contributing to a deeper understanding of the thermal history of the universe and the properties of DM.

\clearpage





\clearpage

\section*{Methods}

\subsection{Simulations of the 1D power spectrum of the 21-cm forest.}
The differential brightness of the observed 21-cm absorption signal relative to the brightness temperature of the background radiation at a specific direction $\hat{\mathbf{s}}$ and redshift $z$ is
\begin{equation}
\delta T_{\mathrm{b}}(\hat{\mathbf{s}}, \nu) \approx \frac{T_{\mathrm{S}}(\hat{\mathbf{s}}, z) -T_\gamma\left(\hat{\mathbf{s}}, \nu_0, z\right)}{1+z} \tau_{\nu_0}(\hat{\mathbf{s}}, z),
\end{equation}
where $\nu_0 =$ 1420.4~MHz is the rest frame frequency of 21-cm photons, $T_{\mathrm{S}}$ is the spin temperature of HI gas, $\tau_{\nu_0}$ is the 21-cm optical depth. In terms of average gas properties within each voxel, the 21-cm optical depth can be written as~\citep{Furlanetto:2006dt,Madau:1996cs,field1959attempt}
\begin{equation}
\begin{aligned}
\tau_{\nu_0}(\hat{\mathbf{s}}, z) \approx 0.0085[1+\delta(\hat{\mathbf{s}}, z)](1+z)^{3 / 2}\left[\frac{x_{\mathrm{HI}}(\hat{\mathbf{s}}, z)}{T_{\mathrm{S}}(\hat{\mathbf{s}} , z)}\right]
\left[\frac{H(z) /(1+z)}{\mathrm{d} v_{\|} / \mathrm{d} r_{\|}}\right] \left(\frac{ \Omega_{\mathrm{b}} h^2}{0.022}\right)\left(\frac{0.14}{\Omega_{\mathrm{m}} h^2}\right),
\end{aligned}
\end{equation}
where $\delta(\hat{\mathbf{s}}, z)$, $x_{\mathrm{HI}}(\hat{\mathbf{s}}, z)$ and $H(z)$ are the gas overdensity, the neutral fraction of hydrogen and the Hubble parameter, respectively. ${\mathrm{d} v_{\|} / \mathrm{d} r_{\|}}$ is the inherent velocity projected along the line of sight gradient and we ignore this term in the simulation. Additionally, doppler shifts due to any peculiar velocities are also ignored.

For the spin temperature, we assume that it is coupled to $T_{\rm K}$ ($T_{\rm S} \approx T_{\rm K}$), due to the Wouthuysen-Field effect~\citep{wouthuysen1952excitation,field1959spin}. The gas temperature in each voxel is set by the thermal history of the early universe and its proximity to halos. Here, our focus is on neutral regions, where X-ray penetration primarily heats the gas, and subsequently affects the 21-cm signals. When we consider an unheated IGM ($f_{\rm X} = 0$), the temperature of gas outside the virial radius is mainly determined by the cosmic expansion. The 21-cm forest signal is more prominent on smaller scales, and thus our work prioritizes the small-scale features. As described in Ref.~\citep{Pritchard:2006sq}, while X-ray heating introduces temperature variations on large scales due to the spatial distribution of X-ray sources, the long mean free path of X-ray photons ensures that, on scales of $\sim 10$ Mpc, the background temperature can be reasonably approximated as uniform. Additionally, we also account for the temperature inhomogeneities within halos and in their surrounding regions: inside halos, the temperature is set to the virial temperature, while outside halos, it depends on X-ray heating. This approach ensures that the influence of small-scale variations is included in our modeling. In addition, our work primarily considers the upper limit of halos with a virial temperature of $10^4 {\rm~K}$. Below this threshold, it is difficult for stars and galaxies to form within the halo. Thus, these halos do not contain X-ray sources, allowing us to assume a relatively uniform distribution of X-ray sources, leading to uniform heating of IGM. While spatial variations in heating may exist in reality~\citep{Park:2018ljd}, this approximation serves as a starting point for this work, which prioritizes the development of an inference method that can be extended to incorporate more complex physical scenarios in future work. For the brightness temperature of the background radiation $T_{\gamma}$, our assumption is the same as that in Ref.~\citep{Shao:2023agv}, which is related to the flux density of the background source $S_{150}$.

We employ the multi-scale modeling approach proposed in Ref.~\citep{Shao:2023agv} to simulate the 21-cm forest signal. Specifically, we use {\tt 21cmFAST}~\citep{Mesinger:2010ne} to model the ionization field~\citep{yue2012reionization,dayal2017reionization} and density field~\citep{Cooray:2002dia,Zentner:2006vw,Press:1973iz,Smith:2011ev} on large scales in a ($1$ Gpc)$^3$ box with $500^3$ grids. Subsequently, we identify neutral regions within the large-scale box to simulate the 21-cm signal, where each grid with a length of $2$~Mpc is further divided into $500^3$ voxels by small-scale simulations. The initial conditions for the grids are derived from large-scale density field simulations. Since low-mass halos are the main contributors to the 21-cm forest signal~\citep{Shimabukuro:2014ava,Shimabukuro:2019gzu,Shimabukuro:2022afm,Shimabukuro:2020tbs,Xu:2021zkf}, we calculate their number and distribution within small-scale grid based on the conditional mass function. The density of each voxel is determined by its nearest halo, following the Navarro-Frenk-White profile~\citep{Navarro:1996gj} within the virial radius and the infall model profile~\citep{Barkana:2002bm} outside of it.
Throughout this work, we adopted a set of cosmological parameters consistent with Planck 2018 results~\citep{Planck:2018vyg}: $\Omega_{\mathrm{m}}= 0.3153$, $\Omega_{\mathrm{b}}h^{ 2}=0.02236$, $\Omega_{\Lambda}= 0.6847$, $h= 0.6736$ and $\sigma _{8}= 0.8111$. Here, $\Omega_{\mathrm{m}}$ is the matter density, $\Omega_{\mathrm{b}}$ is the baryonic density, $\Omega_{\Lambda}$ is the dark energy density, $h$ is the dimensionless hubble parameter, and $\sigma_{8}$ is the matter fluctuation amplitude.

Observational uncertainties in the 21-cm forest include thermal noise, sample variance, contaminating spectral structure of foreground sources in the chromatic side lobes, and bandpass calibration errors~\citep{Shao:2023agv}. Regarding the error consideration, we only consider the thermal noise of the interferometer array and the sample variance in the 1D power spectrum measurement as in Ref.~\citep{Shao:2023agv}. The sample variance is generated by the simulation process and does not need to be added separately. When measuring individual absorption lines directly, the noise flux density averaged over the two polarizations can be expressed as~\citep{thompson2017interferometry}
\begin{equation}
\delta S^{\mathrm{N}} \approx \frac{2 k_{\mathrm{B}} T_{\mathrm{sys}}}{A_{\mathrm{eff}} \sqrt{2 \delta \nu \delta t}},
\end{equation}
where $A_{\mathrm{eff}}$ is the effective collection area of the telescope, $T_{\mathrm{sys}}$ is the system temperature, $\delta \nu$ is the channel width, $\delta t$ is the integration time. The corresponding thermal noise temperature is
\begin{equation}
\delta T^{\mathrm{N}}=\delta S^{\mathrm{N}}\left(\frac{\lambda_z^2}{2 k_{\mathrm{B}} \Omega}\right) \approx \frac{\lambda_z^2 T_{\mathrm{sys }}}{A_{\mathrm{eff}} \Omega \sqrt{2 \delta \nu \delta t}},
\end{equation}
where $\lambda_z$ is the observation wavelength, $\Omega = \pi(\theta / 2)^{2}$ is the solid angle of the telescope beam, where $\theta$ = $1.22 \lambda_z / D$ is the angular resolution, $D$ is the longest baseline of a radio telescope or array. For SKA1-LOW, we use $A_{\mathrm{eff}}/T_{\mathrm{sys }} =$ $800$~$\rm {m^{2}/K}$~\citep{Braun:2019gdo,Acedo:2020lve}. For SKA2-LOW, we use $A_{\mathrm{eff}}/T_{\mathrm{sys }} =$ $4000$~$\rm {m^{2}/K}$~\citep{Braun:2019gdo,SKAO}. For these two arrays, we assume $D =$ $65$~km, and the integration time $\delta t$ is $100$~h and $200$~h, respectively.

In Supplementary Fig.~\ref{sfig9}, we calculated the brightness temperature for different cases. Direct measurement of individual absorption lines is easily hampered by early X-ray heating. To improve the sensitivity of detecting the 21-cm forest signal and reveal the clustering properties of the absorption lines to distinguish the effect on the 21-cm signal between the heating effect and WDM model, we followed the algorithm in Ref.~\citep{Shao:2023agv} and calculated 1D power spectrum of brightness temperature under the assumption of high redshift background sources~\citep{Thyagarajan:2020nch}. The brightness temperature $\delta T_{\mathrm{b}}(\hat{\mathbf{s}}, \nu)$ as a function of the observation frequency $\nu$ can be equivalently used to represent the line-of-sight distance $r_z$ as $T'_{\mathrm{b}}(\hat{\mathbf{s}}, r_z)$. The Fourier transform of $T'_{\mathrm{b}}
(\hat{\mathbf{s}}, r_z)$ is
\begin{equation}
\delta \widetilde{T'}\left(\hat{\mathbf{s}}, k_{\|}\right)=\int \delta T'_{\mathrm{b}}\left(\hat{\mathbf{ s}}, r_z\right) \mathrm{e}^{-i k_{\|} r_z} \mathrm{~d} r_z.
\end{equation}
The 1D power spectrum along the line of sight is defined as
\begin{equation}
P\left(\hat{\mathbf{s}}, k_{\|}\right)=\left|\delta \widetilde{T}^{\prime}\left(\hat{\mathbf{s}} , k_{\|}\right)\right|^2\left(\frac{1}{\Delta r_z}\right).
\end{equation}
Here $1/{\Delta r_z}$ is the normalization factor, where ${\Delta r_z}$ is the line of sight length considered. For a reasonable number $\mathcal{O}$~(10) high-$z$ background sources~\citep{Niu:2024eyf}, the expected value of the 1D power spectrum is obtained by incoherently averaging over neutral patches on the line of sight penetrating various environments, i.e. $P\left(k_{\|}\right) \equiv \left \langle   P\left(\hat{\mathbf{s}}, k_{\|}\right)\right \rangle$. The background sources are assumed to be located outside the other end of the simulation volume, with lines of sight randomly directed through the simulation, each crossing multiple neutral regions. This setup ensures a realistic representation of the spatial distribution of neutral patches and the effects of varying local density. In the rest of this work, we abbreviate $k_{\|}$ to $k$ because here we are always interested in $k$ along the line of sight. We calculated the 1D power spectrum as presented in Supplementary Fig.~\ref{sfig2} of the 21-cm forest corresponding to different conditions. The dotted line indicates the thermal noise of SKA2-LOW when the integration time is $200$ h. It can be seen that utilizing the 1D power spectrum, the 21-cm forest signals can be detected.

Within the large-scale simulation box, we randomly selected 10 neutral hydrogen regions corresponding to 10 background radio sources at redshift $z=9$ with flux densities $S_{150}=10~{\rm mJy}$. Each region contains 5 adjacent grid cells along the line of sight for detailed small-scale modeling. By tessellating these 2 Mpc grid units, we established a composite grid system spanning 2 Mpc $\times$ 2 Mpc $\times$ 10 Mpc, containing $500 \times 500 \times 2500$ voxels (i.e., $500 \times 500$ independent line-of-sight paths, each penetrating a 10 Mpc-depth structure). This configuration preserves the 10 Mpc-scale correlations derived from the {\tt 21cmFAST} simulations at large scales while introducing negligible artifacts in the 1D small-scale power spectrum along the line of sight. To mitigate edge effects caused by excluding external DM halos in our simulation, we implemented boundary truncation in two transverse dimensions: removing peripheral layers from the original $500 \times 500$ line array perpendicular to the line-of-sight direction, retaining only the central $300 \times 300 = 90000$ valid line-of-sight paths. For each background source, we assumed that 10 distinct 10 Mpc neutral hydrogen segments could be extracted. Given potential intervening ionized bubbles between neutral segments, we adopted a conservative estimate of 200 Mpc total line-of-sight length, equivalent to a redshift span $\Delta z \approx 0.8$ --- a range consistent with practical observational constraints. To effectively suppress thermal noise, we simulated 90000 power spectra per background source. For each source, these power spectra were partitioned into 9000 groups, each containing 10 line-of-sight paths. All 10 paths within a group originate from identical simulation parameters, thereby representing 10 neutral hydrogen segments in a single background source spectrum. Finally, through incoherent averaging of 100 power spectra (10 sources $\times$ 10 segments), we generated 9000 synthetic 1D power spectra.

However, the infall model, suitable for high-density areas with numerous halos, tends to overestimate the density field, requiring normalization of the field to match the initial local density of {\tt 21cmFAST} outputs. The normalization technique of Ref.~\citep{Shao:2023agv}, which involves multiplying each voxel by a factor, alters the amplitude of the 1D power spectrum of the 21-cm forest and leads to a crossover in the 1D power spectrum across different DM models (See Fig.~$3$ in Ref.~\citep{Shao:2023agv}). To address this, we employ a normalization method that does not affect the amplitude of the 1D power spectrum, adjusting the density field by adding or subtracting a number. Furthermore, to maintain the probability density distribution of the density field, we adjust the lower limit of halo mass to $10^{5} M_{\odot}$ in the simulation, corresponding to the Jeans mass~\citep{Ripamonti:2006gr} at the redshift of interest, which also mitigates anomalies in the 1D power spectrum on large scales. These adjustments to the density field and halo mass lower limit modify the structure of the 1D power spectrum at different scales, thereby changing the direction of parameter degeneracies. Figure $6$ in Ref.~\citep{Shao:2023agv} shows that the degeneracy direction between $m_{\rm WDM}$ and $T_{\rm K}$ is negatively correlated. As we previously discussed in Supplementary Fig.~\ref{sfig7}, different scales affect the degeneracy direction of the two parameters differently. In this work, the improvements we made have altered the relationship of the 1D power spectrum on large scales for different DM models. Consequently, in Fig.~\ref{fig conter}, the degeneracy direction between $m_{\rm WDM}$ and $T_{\rm K}$ becomes positively correlated. Meanwhile, using SKA2-LOW with an integration time of $100$ h is insufficient to observe the 21-cm forest signal. As a result, we increased the integration time to $200$ h.

\subsection{Dataset settings.}
We finally obtained the simulated correspondence between parameters ($m_{\rm WDM}$, $T_{\rm K}$) and the 1D power spectrum in the presence of thermal noise. Specifically, we simulated $7$ simulation regions with DM particle masses ranging from $3$ keV to $9$ keV at equal intervals ($1$ keV). For the gas temperature, we consider two situations. When the heating effect is weak ($f_{\rm X}\approx  0.1$), the IGM will be heated to about $60$~K at $z=9$. Therefore, we choose equal intervals ($5$~K) between $40$~K and $80$~K. When the heating effect is strong ($f_{\rm X}\approx  1$), the IGM will be heated to about $600$~K at $z=9$. Therefore, we choose equal intervals ($100$~K) between $400$~K and $800$~K. For each set of parameters, out of 9000 power spectra, we take 300 as the testing set and use the remaining 8700 for training. In addition, we also generated some data sets of other parameters for validation and testing sets as presented in Supplementary Fig.~\ref{sfig10}. It should be noted that this method is not limited to the above parameter range and is also applicable to other parameter ranges. Although LHS~\citep{mckay2000comparison,Schmit:2017pho} could reduce the number of sampling points per dimension, the computational cost of varying $T_{\rm{K}}$ is relatively small once $m_{\rm WDM}$ is fixed, making LHS less beneficial in this context.
Specifically, for the density field simulation, when the mass is 9 keV, it can take up to $1317$ h on $64$ cores (Intel(R) Xeon(R) Gold 6271C CPU @ 2.60GHz). In contrast, the simulation of temperature field can be completed within 1 min. Moreover, given the correlation between $m_{\rm WDM}$ and $T_{\rm{K}}$, this approach would be less effective.  For future studies involving more parameters, exploring higher-dimensional spaces and considering methods like LHS to improve efficiency are worthwhile directions for further investigation. 
However, in studies of the 21-cm forest, the dominant factors influencing the signal arise from DM properties governing the density field and heating effects modulating the temperature field. Cosmological parameters exhibit minimal observational impact during the EoR, while astrophysical parameters linked to reionization processes remain subject to significant uncertainties. We therefore prioritize these two pivotal parameters that critically shape the 21-cm forest signal, as they encapsulate the dominant physical mechanisms relevant to near-future investigations.


To fit the input parameters to the deep learning model, we first need to normalize the 1D power spectrum and parameters before training and testing the network. For the parameters, we perform normalization as the formula
\begin{equation}
\tilde{\xi }= 2 \frac{\xi - \xi_{min}}{\xi_{\max} -\xi_{\min}} -1,
\end{equation}
where $\xi$ represents the original parameter ($m_{\rm{WDM}}$ or $T_{\rm{K}}$), $\xi_{\max}$ is the maximum value of $\xi$, and $\xi_{\min}$ is the minimum value of $\xi$. Due to the significantly higher values of the power spectrum near the halo center, we used logarithmic normalization. This method helps handle the wide range of values and the outliers, ensuring more uniform analysis across different scales. We perform logarithmic normalization on the 1D power spectrum as the formula
\begin{equation}
\tilde{P} (k)= 2 \frac{\log P(k) - \log P_{\min}(k)}{\log P_{\max}(k) -\log P_{\min}(k)} -1,
\end{equation}
where $P(k)$ represents the value of the 1D power spectrum at a specific wavenumber $k$ after binning, $\log P_{\max}(k)$ is the logarithm of the maximum 1D power spectrum value at $k$, and $\log P_{\min}(k)$ is the logarithm of the minimum 1D power spectrum value at $k$. This approach helps balance the contributions of each scale and improves the stability and efficiency of network training. The data set generated and processed by the above method are presented in Supplementary Fig.~\ref{sfig10}, which not only provides a solid foundation for training and testing our deep learning model but also ensures the reliability and scientific significance of the research results. Additionally, we conducted a test on two completely random points, and the results, as shown in Supplementary Fig.~\ref{sfig11}, validate the model. 

\subsection{The principle of the normalizing flow.}
Based on the above simulation, we propose a parameter inference method. The basic idea is to generate a large number of simulated data sets with relevant parameters and use these data sets to train a neural network, specifically an NF model~\citep{rezende2015variational} to approximate the posterior distribution. Compared to traditional generative models such as generative adversarial networks~\citep{goodfellow2014generative} and variational autoencoders~\citep{DBLP:journals/corr/KingmaW13}, which often suffer from challenges like mode collapse, posterior collapse, and imprecise probability estimation~\citep{bond2021deep}, GNF provides a significant advantage by directly learning the exact likelihood function, which has been proven to enable high-quality data generation~\citep{Hassan:2021ymv}. 
In addition, NFs have demonstrated their gravitational wave parameter estimation capabilities by enabling real-time inference in high-dimensional spaces (up to 17 dimensions)~\citep{Dax:2024mcn}, with scalability potential to higher dimensionality. Recent methodological advances reveal that emerging alternatives like diffusion models~\citep{DBLP:journals/corr/abs-2006-11239} achieve superior sample generation quality while addressing aforementioned limitations, yet their multi-step iterative processes incur substantial computational latency. By contrast, the NF framework maintains implementation practicality --- a critical advantage requiring systematic evaluation when deploying such models in time-sensitive astrophysical applications.
By leveraging these simulated data sets, the trained network can quickly generate new posterior samples when observational data is obtained and tested. This eliminates the need to generate the 1D power spectrum at inference time, effectively amortizing the computational cost of training across all future detections. The general method for building such models is called neural posterior estimation (NPE)~\citep{papamakarios2016fast}.


In more detail, the NF is a method of probabilistic modeling using deep learning technology. Here, we use the notation of the INF for explanation. The differences between the INF and GNF, as shown in the Fig.~\ref{fig NSF}, are merely the conditions and outputs. The core idea of NF is to construct a bijective transformation $f_P$ based on the 1D power spectrum $P$, thereby obtaining the mapping relationship between a simple base distribution (e.g., a normal distribution) and a complex posterior distribution. It is a type of reversible neural network characterized by the ability to accurately calculate the Jacobian determinant of the probability density function through network transformation. The NF can be mathematically expressed as~\citep{rezende2015variational}
\begin{equation}\label{eq1}
\mathcal{Q}(\xi|P)=\pi\left(f_P^{-1}(\xi)\right)\left|\operatorname{det} J_{f_P}^{-1}\right|,
\end{equation}
where $\mathcal{Q}(\xi|P)$ represents the approximate posterior distribution exported by the NF, $\pi(u)$ is typically chosen to be a standard multivariate normal distribution for ease of sampling and density evaluation, and $f_P$ represents the transformation applied to the data. This transformation can be applied iteratively to construct complex densities:
\begin{equation}
f_P(u)=f_{P,~N}\circ f_{P,~N-1}\circ \dots \circ f_{P,~1}(u),
\end{equation}
where each $f_{P,~i}(u)$ represents a block of the NF. This iterative approach is central to NPE, where the goal is to train a parameter conditional distribution that approximates the true posterior distribution. This task translates into an optimization problem aimed at minimizing the expected Kullback-Leibler divergence (KLD) between the true and approximate posterior distributions~\citep{10.1214/aop/1176996454}:
\begin{equation}
{\rm KLD}(\mathcal{P}\parallel \mathcal{Q})=\int_{-\infty }^{+\infty } \mathcal{P}{(\xi)}\log \frac{\mathcal{P}{(\xi)} }{\mathcal{Q}{(\xi)} } \mathrm{d}\xi,
\end{equation}
where $\mathcal{P}{(\xi)}$ represents the true posterior distribution and $\mathcal{Q}{(\xi)}$ represents the approximate posterior distribution.

The training of the NF involves minimizing the loss function, which is the expected value of the cross-entropy between the true and model distributions~\citep{papamakarios2019sequential}:
\begin{equation}
L=\mathbb{E}_{\mathcal{P}(P)}\left[\operatorname{KLD}\left(\mathcal{P}(\xi \mspace{-5mu} \mid \mspace{-5mu} P) \| \mathcal{Q}(\xi \mspace{-5mu} \mid \mspace{-5mu} P)\right)\right].
\end{equation}
For a minibatch of training data of size $N$, this can be approximated as
\begin{equation}
L \approx-\frac{1}{N} \sum_{i=1}^N \log \mathcal{Q}\left(\xi^{(i)} \mspace{-5mu} \mid \mspace{-5mu} P^{(i)}\right).
\end{equation}

Our NF implementation is based on the neural spline flow (NSF)~\citep{durkan2019neural}, specifically a rational-quadratic neural spline flow with autoregressive layers, RQ-NSF (AR). Each flow is connected by a series of autoregressive layers that share parameters. Specifically, the network uses the conditional and partially vectorized output from the previous flow (when $i$ is equal to $0$, this is the vector sampled from a standard distribution) as the input of a multilayer perceptron (MLP), and the output of the MLP is used to parameterize the RQ-NSF (AR). These parameters are defined as the width and height of $K$ bins, which ultimately form $K+1$ knots $\left \{ \left ( x^{(i)},y^{(i)} \right ) \right \} _ {i=0}^{K}$. The left and right endpoints are $(-B, -B)$ and $(B, B)$, respectively, and there are $K-1$ derivatives of knots $\left \{ \dot{y} ^{(i)} \right \} _{i=1}^{K-1}$, where the derivatives of the left and right endpoints are set to 1. Therefore, the data is divided into $K$ bins from $-B$ to $B$. Each stream is a map that maps $-B$ to $B$ and then $-B$ to $B$ again, so the y-axis is also divided into $K$ bins from $-B$ to $B$. Figure 1 in Ref.~\citep{durkan2019neural} shows the mapping relationship of the coupling transformation when $K$ is $10$. 

\subsection{Neural network architecture.}
NFs, unlike Bayesian methods, do not require an explicit definition of the likelihood function. They employ a likelihood-free approach, making it suitable for handling non-Gaussian data, where deriving an analytical likelihood is either computationally expensive or infeasible. Since the scale of the 21-cm forest data is extremely small, generating the 1D power spectra of the 21-cm forest corresponding to a set of parameter points requires even up to $40000$ core hours. This will make it difficult to obtain 1D power spectrum for any parameter, potentially leading to incorrect estimates of the posterior distribution as presented in Supplementary Fig.~\ref{sfig1}. Therefore, our method consists of a GNF for generating the 1D power spectrum of arbitrary parameters and an INF for parameter inference. The primary difference between these two is the conditioning: the GNF conditions on the parameters to generate power spectra, while the INF conditions on the power spectra to infer parameters. All the NF networks were trained using the ``Xavier''~\citep{glorot2010understanding} initialization for network parameters and the AdamW optimizer~\citep{loshchilov2018decoupled} on a single NVIDIA GeForce RTX A6000 GPU with 48 GB of memory.

The GNF is used to generate a 1D power spectrum corresponding to all parameters in the parameter space. Interpolation is a common method for filling in missing data and can serve as a supplementary approach for generating the 1D power spectra. However, as shown in Supplementary Fig.~\ref{sfig12}, due to the inherent cosmic variance and the correlations between different scales $k$, interpolation often results in lower $P(k)$ and less smooth transitions in the interpolated regions of parameter space. This highlights the limitations of interpolation methods, as they rely heavily on the assumption of smooth parameter variation, which is not always valid in the presence of cosmic variance. In contrast, the GNF model does not merely act as an interpolation tool but learns the underlying statistical relationships between parameters and the 1D power spectra, enabling it to reproduce both the smooth and stochastic features of the data. However, as is common in deep learning models, the GNF model may produce less reliable extrapolations in regions of parameter space that are underrepresented in the training data. For example, while the model performs well at high $T_{\rm K}$, it may yield unphysical results for large-scale modes at low $T_{\rm K}$. This limitation can be addressed by expanding the range of the training dataset to better cover the parameter space. Specifically, the GNF takes a parameter vector as input and finally outputs an estimate of the 1D power spectrum of the 21-cm forest. The training set is composed of $8700$ black points from the data, as shown in Supplementary Fig.~\ref{sfig10}, the blue point data as the validation set, and the red point as the testing set. For GNFs, to prevent overfitting and achieve convergence, we employed early stopping with a patience of 15 epochs and utilized $L_1$ regularization and $L_2$ regularization~\citep{hoerl1970ridge}. A summary of key hyperparameters for GNF is provided in Supplementary Table~\ref{stable1}.

The INF is designed to perform parameter inference based on data generated by the GNF. The architecture of the INF is similar to that of the GNF. Specifically, the INF receives the 1D power spectra generated by the GNF, conditioned on the 1D power spectra $P(k)$, and maps the corresponding parameters $\theta$ onto a standard distribution. By utilizing the bijective transformation property of the INF, this process enables the reverse prediction of the model parameters $\theta$. If the GNF can accurately capture the characteristics of the 1D power spectrum, then using the sampling points from the GNF within the prior range as the input of the INF, the INF can provide appropriate inference results for the simulation data. To prevent overfitting, the INF is trained on dynamically simulated data generated by the GNF in an iterative scheme. Specifically, during each training step of the INF, the GNF generates new samples based on the current training progress. These samples are used to update the INF's parameters, and this iterative process continues until the loss function stabilizes, indicating that the training has converged. For this work, we use a batch size of $64$, and the GNF generates $128$ samples per batch during training. A summary of key hyperparameters for INF is provided in Supplementary Table~\ref{stable1}. The remaining $300$ black points in the data are used as the testing set as presented in Supplementary Fig.~\ref{sfig10}.

\subsection{Model evaluation method.}
We use the coefficient of determination $R^2$ to evaluate whether the mean and standard deviation of the 1D power spectra generated by the GNF are reasonable. Its expression is:
\begin{equation}
R^2 = 1- \frac{\sum (y_{\rm pred}-y)^2}{\sum (y-\bar{y} )^2},
\end{equation}
where $y_{\rm pred}$ and $y$ represent the mean or standard deviation of the 1D power spectra generated by the GNF and the corresponding values from the simulated 1D power spectra across different scales, respectively. $\bar{y} $ represents the overall mean or standard deviation of the simulated generated 1D power spectra on different scales. To  check whether the distributions generated under different conditions and scales are consistent with the simulated situation, we also calculate JSD:
\begin{equation}
{\rm JSD}(\mathcal{P}_{\rm sim}\parallel \mathcal{Q}_{\rm gen} )=\frac{1}{2} {\rm KLD}(\mathcal{P}_{\rm sim} \parallel \frac{\mathcal{P}_{\rm sim}+\mathcal{Q}_{\rm gen}}{2} )+\frac{1}{2} {\rm KLD}(\mathcal{Q}_{\rm gen}\parallel \frac{\mathcal{P}_{\rm sim}+\mathcal{Q}_{\rm gen}}{2} ),
\end{equation}
where $\mathcal{P}_{\rm sim}$ and $\mathcal{Q}_{\rm gen}$ represent the distribution of simulated 1D power spectra and the GNF-generated 1D power spectra on different $k$, respectively. The JSD is a symmetric measure of similarity between two probability distributions. A smaller JSD value indicates a higher degree of similarity between the distributions, effectively addressing the asymmetry issue inherent in KLD.

To evaluate the distribution characteristics of the 1D power spectra of the 21-cm forest at specific scales under different astrophysical conditions, we used the skewness coefficient in statistical analysis as a key indicator to assess its Gaussianity. We calculated the skewness coefficient of the normalized 1D power spectrum at different scales $k$, which is defined as the ratio of the third standardized moment to the cube of the standard deviation:
\begin{equation}
S =\mathbb{E}  \left ( \frac{P(k)-\mu}{\sigma }   \right )^3 ,
\end{equation}
where $P(k)$ represents the normalized 1D power spectrum at different scales $k$, $\mu$ is the mean value of $P(k)$, $\sigma$ is the standard deviation of $P(k)$, and $E$ represents the expected value. If the value of $S$ is close to zero, the distribution can be considered symmetric at that scale and close to a Gaussian distribution. If $S$ is significantly different from zero, it indicates that the distribution has a noticeable skewness.

To quantify the correlation between different scales $k$ of the 1D power spectrum of the 21-cm forest, we use the Spearman correlation coefficient\citep{fieller1957tests}. The Spearman correlation coefficient is a non-parametric measure of rank correlation that assesses the monotonic relationship between two variables. It does not assume a normal distribution of the data, making it more robust for handling non-Gaussian distributions. Given the significant non-Gaussian characteristics of our data, we chose the Spearman correlation coefficient to more accurately capture the correlations between different scales. This approach helps avoid errors that could arise from assuming a Gaussian distribution of the data. If all $n$ ranks are distinct integers, the Spearman correlation coefficient $\rho$ is defined as
\begin{equation}
\rho=1-\frac{6 \sum d_{i}^{2}}{n\left(n^{2}-1\right)},
\end{equation}
where $d_{i}$ is the difference between the ranks of corresponding values of the two variables, and $n$ is the number of observations.

In our analysis, we employ the KS test~\citep{Lopes2011} to assess the fit between the posterior distribution of the INF's output and the theoretical distribution. The KS test compares the empirical distribution derived from the sample data with a specified theoretical distribution by measuring the maximum difference between their CDFs. This difference is known as the KS statistic $d$, which can be expressed as
\begin{equation}
d = \max |F_n(x) - F(x)|,
\end{equation}
where $F_n(x)$ is the empirical CDF of the sample, and $F(x)$ is the theoretical CDF. Theoretically, within a given confidence interval of the posterior distribution, the percentage of actual data should match the confidence level. Therefore, the theoretical CDF should be uniform. To interpret the d value in terms of statistical significance, we calculate a $p$-value using the distribution of the KS statistic under the null hypothesis. This $p$-value represents the probability of observing a KS statistic as extreme as $d$, assuming that the sample distribution is drawn from the theoretical distribution. The CDF of the KS statistic under the null hypothesis is given by
\begin{equation}
Q(d) = 2\sum_{i=1}^{\infty}(-1)^{i-1}e^{-2i^2d^2}.
\end{equation}
The $p$-value is derived from this function and represents the probability of observing a value as extreme as the calculated $d$. A small $p$-value (typically indicated by a significance level of $0.05$) would lead us to reject the null hypothesis, suggesting a significant discrepancy between the empirical and theoretical distributions.





\clearpage
\begin{addendum}
\item[Data Availability]
The datasets for all figures in the text of this work are provided in Supplementary Data 1. Other datasets that support the findings of this work are available from the corresponding author upon reasonable request.

\item[Code Availability]
The code {\tt 21cmFAST} used for large-scale simulation is publicly available at \\ \href{https://github.com/andreimesinger/21cmFAST}{https://github.com/andreimesinger/21cmFAST},
the codes for simulating small-scale structures and 21-cm forest signals are available from the corresponding authors upon reasonable request.
The code for building deep learning networks is publicly available at \href{https://github.com/probabilists/lampe}{https://github.com/probabilists/lampe}, and the codes for training and testing in deep learning are available from the corresponding authors upon reasonable request.

\end{addendum}


\begin{addendum} 
\item[Additional information]

{\bf Correspondence and requests for materials} should be addressed to Xin Zhang (email: zhangxin@mail.neu.edu.cn).

\item
This work was supported by the National SKA Program of China (Grants Nos. 2022SKA0110200 and 2022SKA0110203), the National Natural Science Foundation of China (Grants Nos. 12473001, 11975072, 11875102, and 11835009), and the National 111 Project (Grant No. B16009).

\item[Author contributions] 
Tian-Yang Sun performed most of the computation and wrote the majority of the manuscript. Yue Shao performed part of the computation and wrote part of the manuscript. Yichao Li, Yidong Xu and He Wang wrote part of the manuscript. Xin Zhang and Yidong Xu proposed the study. Xin Zhang led the study and contributed to the manuscript writing. All authors discussed the results and commented on the manuscript.

\item[Competing Interests] 

The authors declare no competing interests.

\end{addendum}

\newpage
\newpage

\newpage

\newpage

\newpage

\newpage
\renewcommand{\thefigure}{A\arabic{figure}}

\begin{sfigure}
\centering
\includegraphics[angle=0, width=16.0cm]{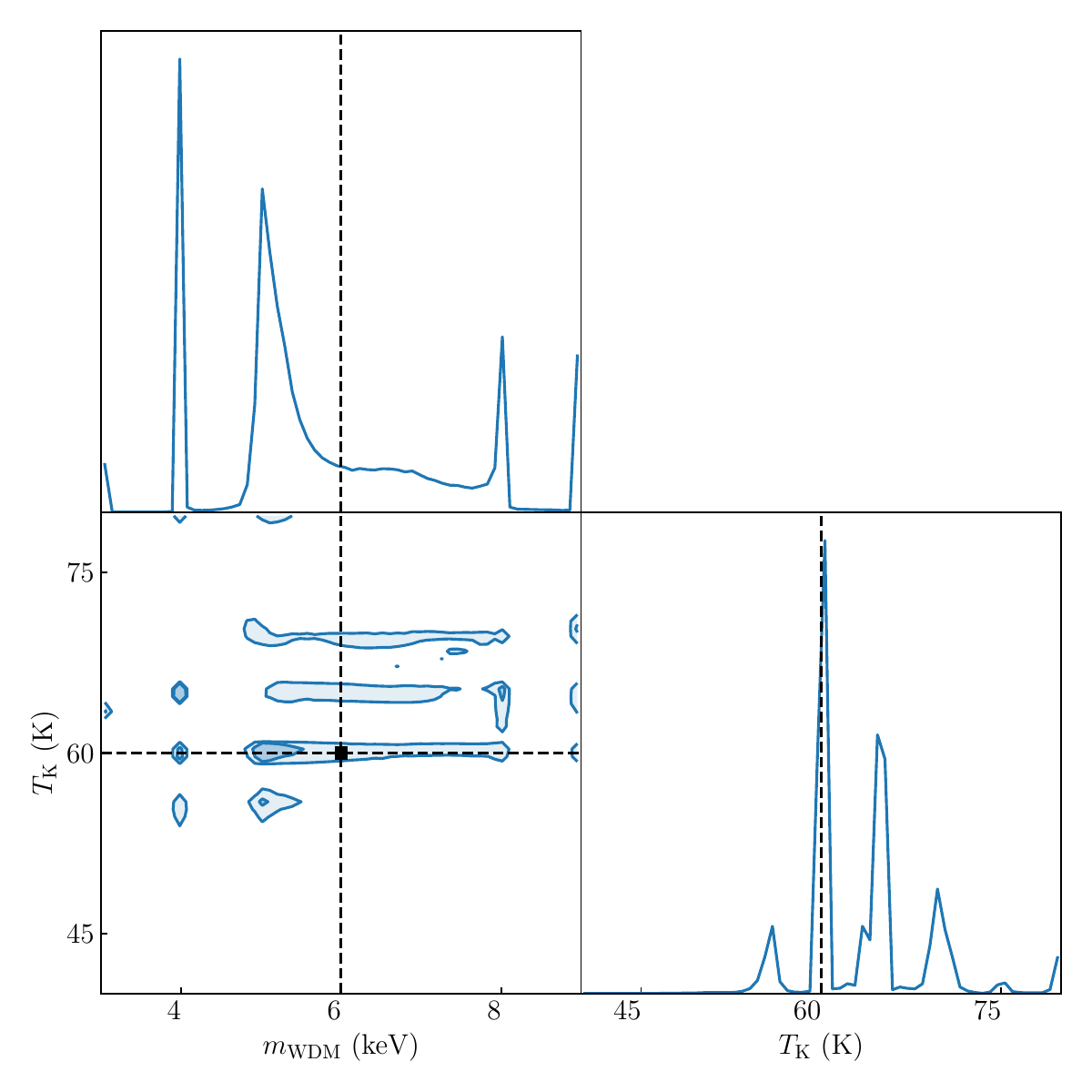}
\caption{\label{sfig1}
  \textbf{The posterior distributions for $m_{\rm WDM}$ and $T_{\rm K}$ using the NF directly without the INF for data enhancement.}
 Due to the missing part of the 1D power spectrum corresponding to the parameters, the NF cannot obtain reasonable parameter inference results. The panel presents the results when $m_{\rm WDM} = 6$ keV and $T_{\rm K} = 60$ K with an integration time of $100$ h using SKA1-LOW. Contours represent $1\sigma$ and $2\sigma$ confidence intervals.
  }
\end{sfigure}

\newpage
\begin{sfigure}
\centering
\includegraphics[angle=0, width=16.0cm]{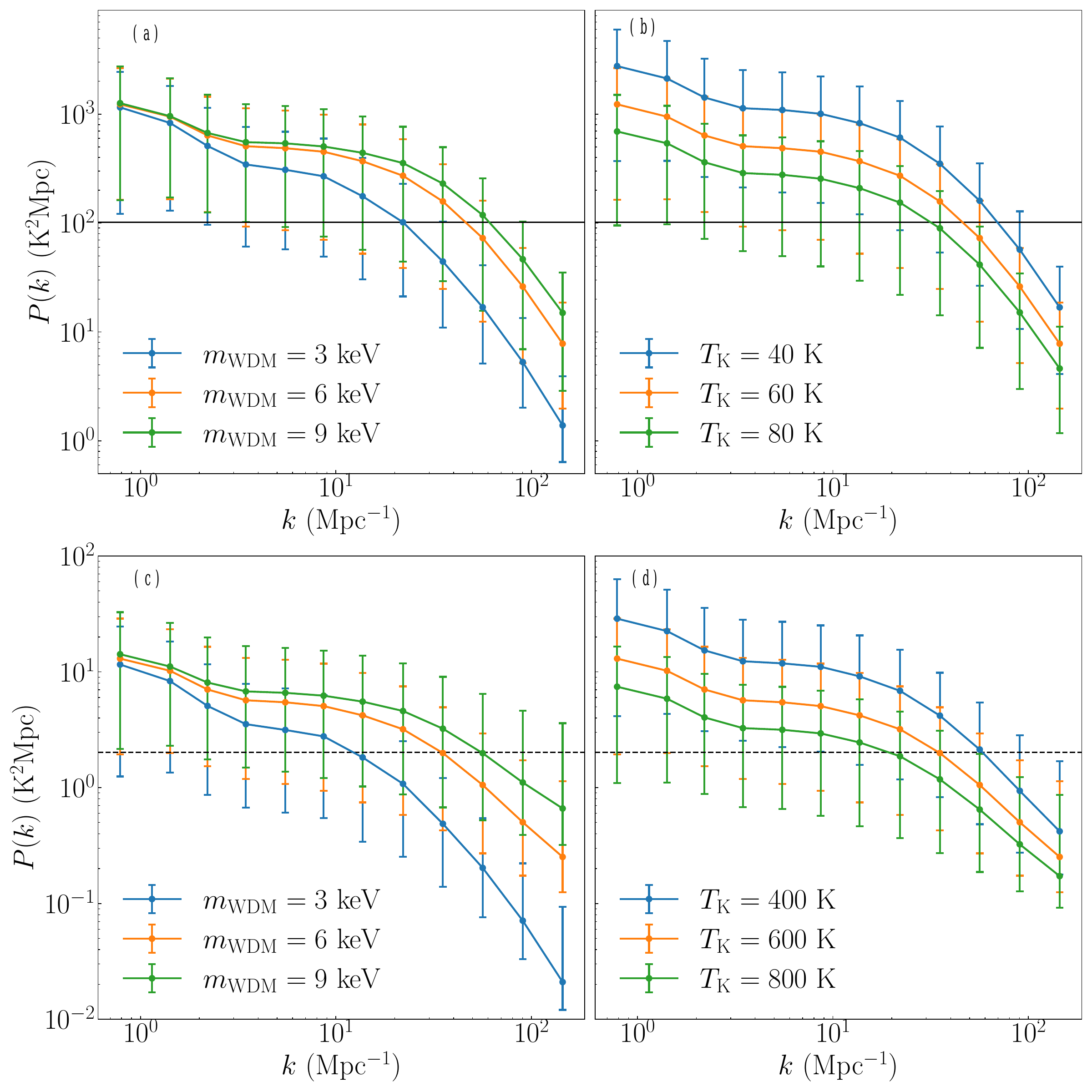}
\caption{\label{sfig2}
  \textbf{Expected 1D power spectrum of the 21-cm forest.}
 A total of $100$ measurements of 10 comoving Mpc length segments along the line-of-sight neutral patch against 10 background light sources with $S_{150} = 10$ mJy yielded the expected value for the 1D power spectrum of 21-cm forest at $z = 9$.
 \textbf{a, c} Depict the 1D power spectra for different $m_{\rm WDM}$ values in the WDM models. The blue, orange, and green curves correspond to $m_{\rm WDM} = 3$ keV, $m_{\rm WDM} = 6$ keV and $m_{\rm WDM} = 9$ keV, respectively. $T_{\rm K}$ in (a) and (c) are 60 K and 600 K, respectively. \textbf{b, d} Depict the 1D power spectra corresponding to the different $T_{\rm K}$. In all cases, $m_{\rm WDM}$ is set to $6$ keV. The blue, orange, and green curves in \textbf{(b)} correspond to $T_{ \rm K} = 40$ K, $T_{\rm K} = 60$ K and $T_{\rm K} = 80$ K. The blue, orange, and green curves in \textbf{(d)} to $T_{\rm K} = 400$ K, $T_{\rm K} = 600$ K and $T_{\rm K} = 800$ K, respectively. The black solid and dashed lines are the thermal noises $P^{\mathrm{N}}$ expected for an integration time of $100$ h using SKA1-LOW and an integration time of $200$ h using SKA2-LOW, respectively. The error bars show sample variances of the 1D power spectra of the 21-cm forest.
  }
\end{sfigure}

\newpage
\begin{sfigure}
\centering
\includegraphics[angle=0, width=16.0cm]{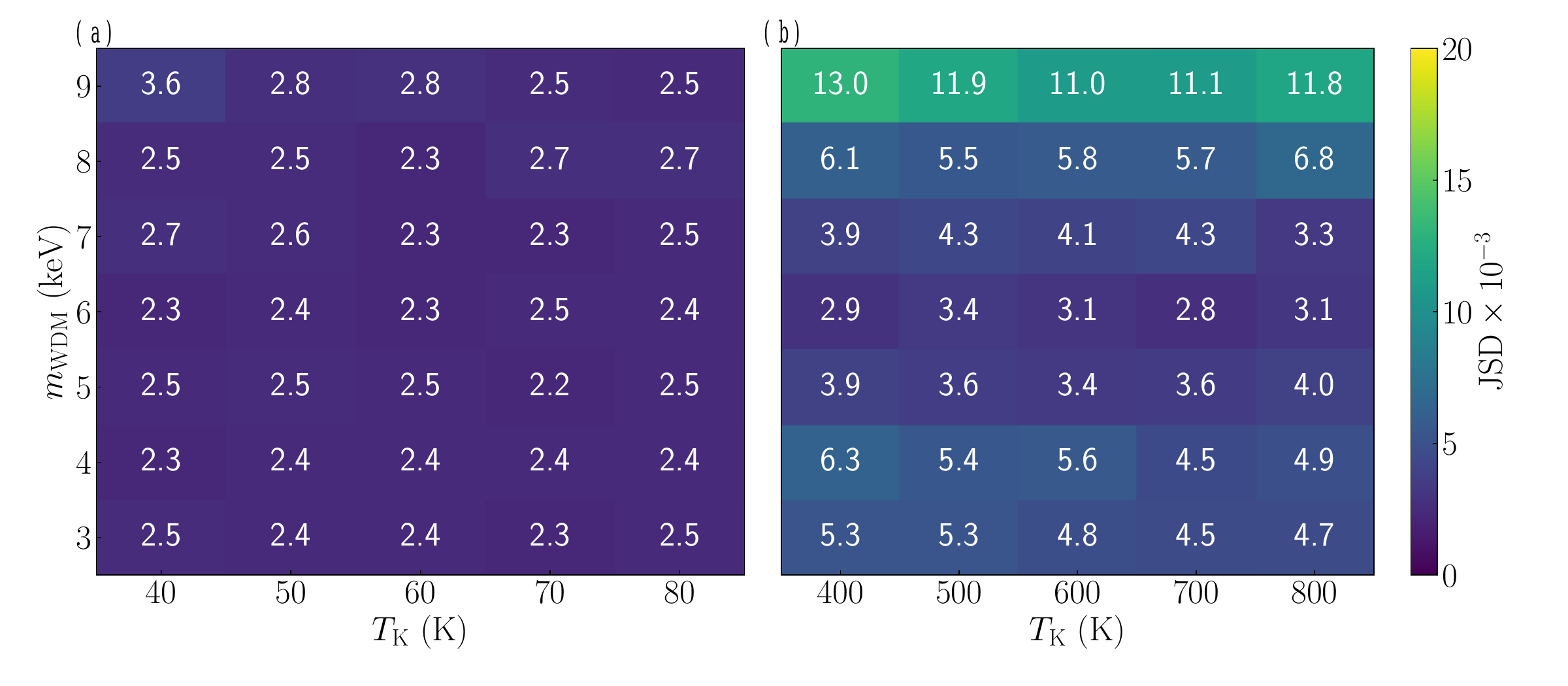}
\caption{\label{sfig3}
  \textbf{JSDs between GNF and simulated power spectra at different $T_{\rm K}$ and $m_{\rm WDM}$.}
  \textbf{a} Depicts the average JSDs of the 1D power spectra generated by different parameters at different $k$ with an SKA1-LOW integration time of $100$ h in the case of a weaker heating effect. \textbf{b} Depicts the average JSDs of the power spectra generated by different parameters at different $k$ with an SKA2-LOW integration time of $200$ h in the case of a stronger heating effect.
  }
\end{sfigure}

\newpage
\begin{sfigure}
\centering
\includegraphics[angle=0, width=16.0cm]{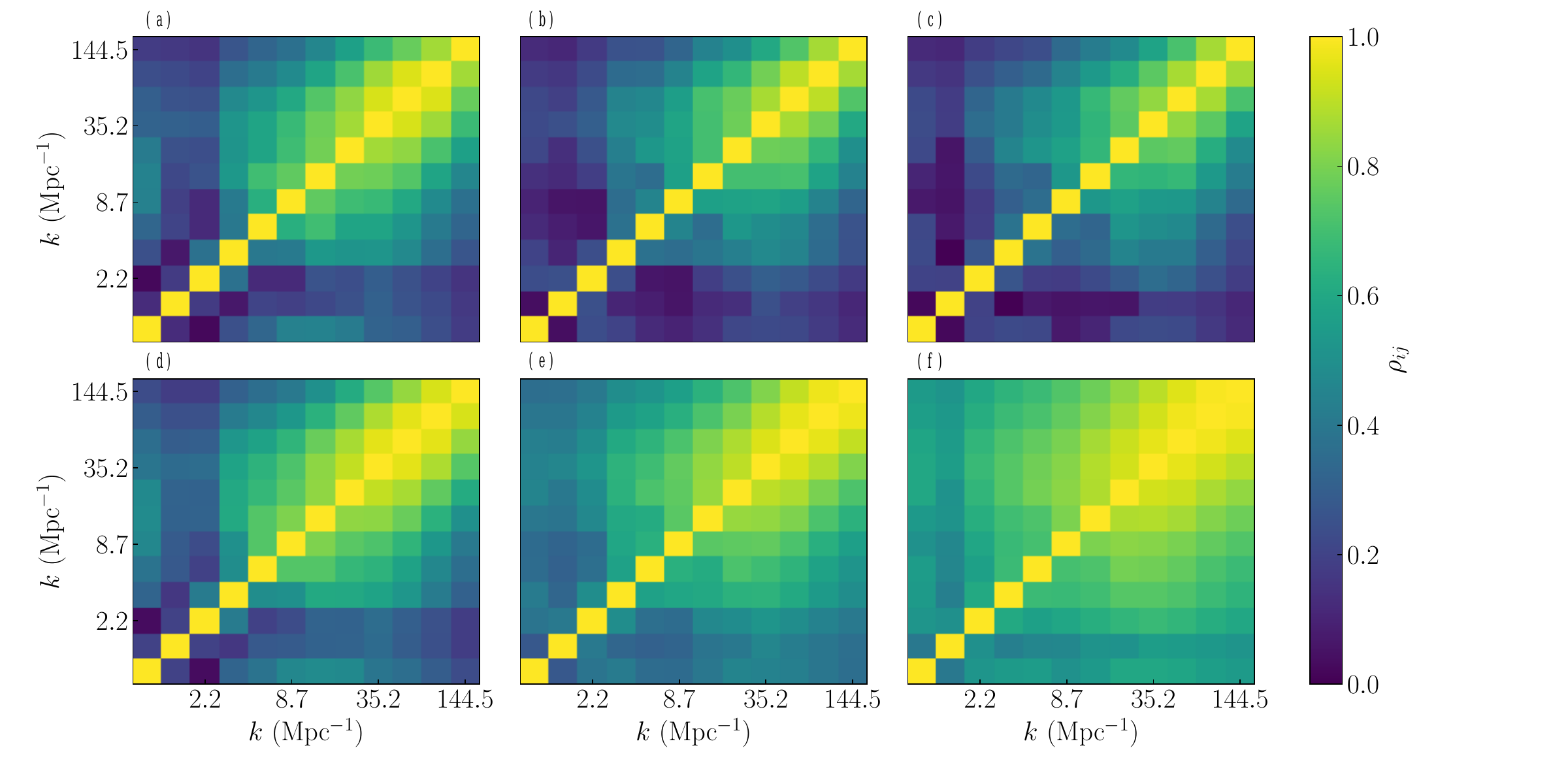}
\caption{\label{sfig4}
  \textbf{The spearman correlation coefficient $\rho_{ij}$ estimated from the 1D power spectrum of 21-cm forest at $z = 9$ without considering thermal noise.}
  \textbf{a - c}, \textbf{d - f} Present $\rho_{ij}$ estimated from the 1D power spectrum of 21-cm forest with weaker ($T_{\rm K} = 60$ K) and stronger ($T_{\rm K} = 600$ K) heating effects, respectively. $\rho_{ij}$ estimated from the 1D power spectrum of 21-cm forest is presented in \textbf{(a, d)}, \textbf{(b, e)} and \textbf{(c, f)} with $m_{\rm WDM} = 3$ keV, $m_{\rm WDM} = 6$ keV and $m_{\rm WDM} = 9$ keV, respectively.
  }
\end{sfigure}

\newpage
\begin{sfigure}
\centering
\includegraphics[angle=0, width=16.0cm]{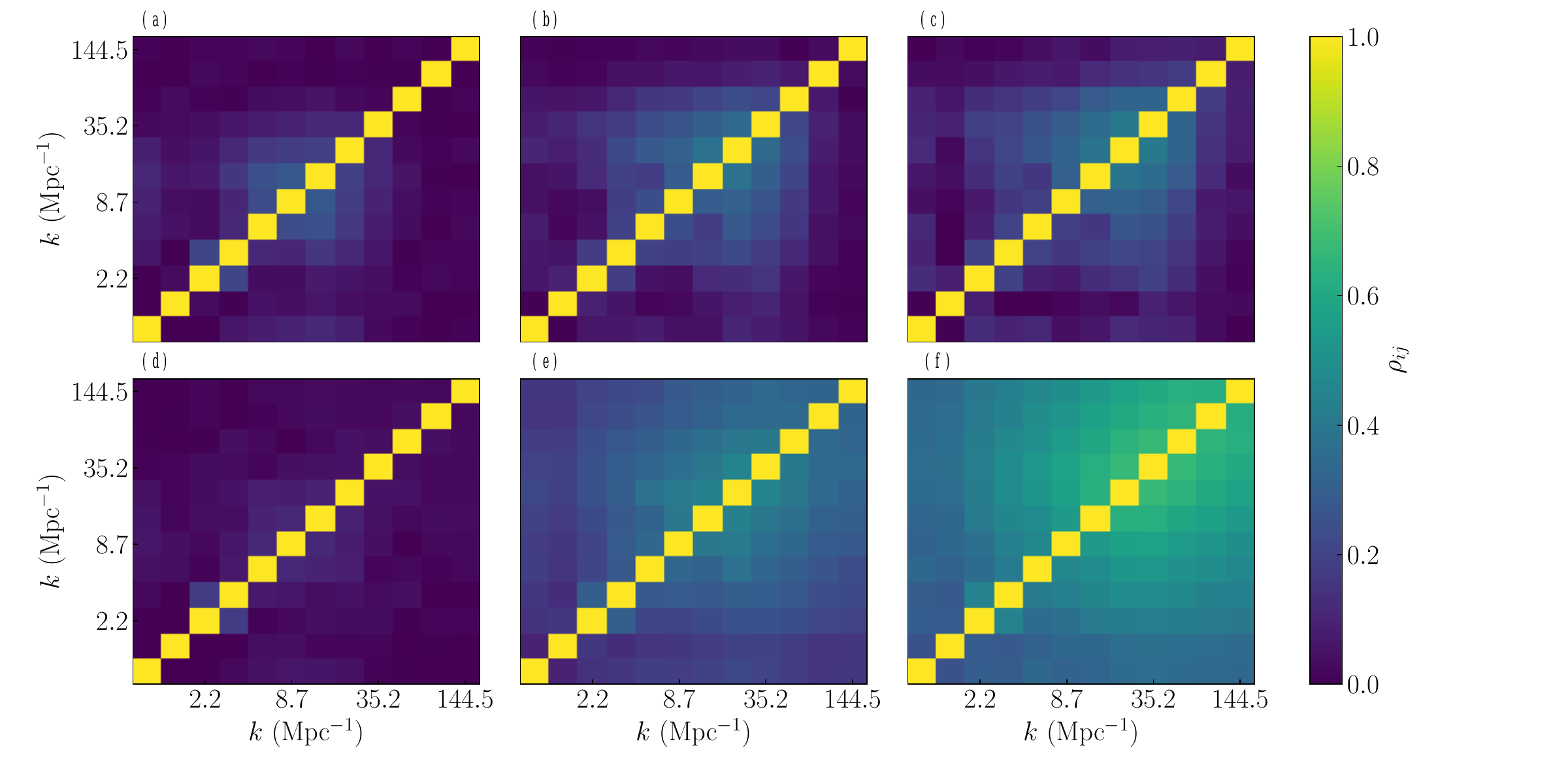}
\caption{\label{sfig5}
  \textbf{The spearman correlation coefficient $\rho_{ij}$ estimated from the 1D power spectrum of 21-cm forest at $z = 9$ with thermal noise.}
  \textbf{a - c}, \textbf{d - f} Present $\rho_{ij}$ estimated from the 1D power spectrum of 21-cm forest with weaker ($T_{\rm K} = 60$ K) and stronger ($T_{\rm K} = 600$ K) heating effects, respectively. $\rho_{ij}$ estimated from the 1D power spectrum of 21-cm forest is presented in \textbf{(a, d)}, \textbf{(b, e)} and \textbf{(c, f)} with $m_{\rm WDM} = 3$ keV, $m_{\rm WDM} = 6$ keV and $m_{\rm WDM} = 9$ keV, respectively.
  }
\end{sfigure}

\newpage
\begin{sfigure}
\centering
\includegraphics[angle=0, width=16.0cm]{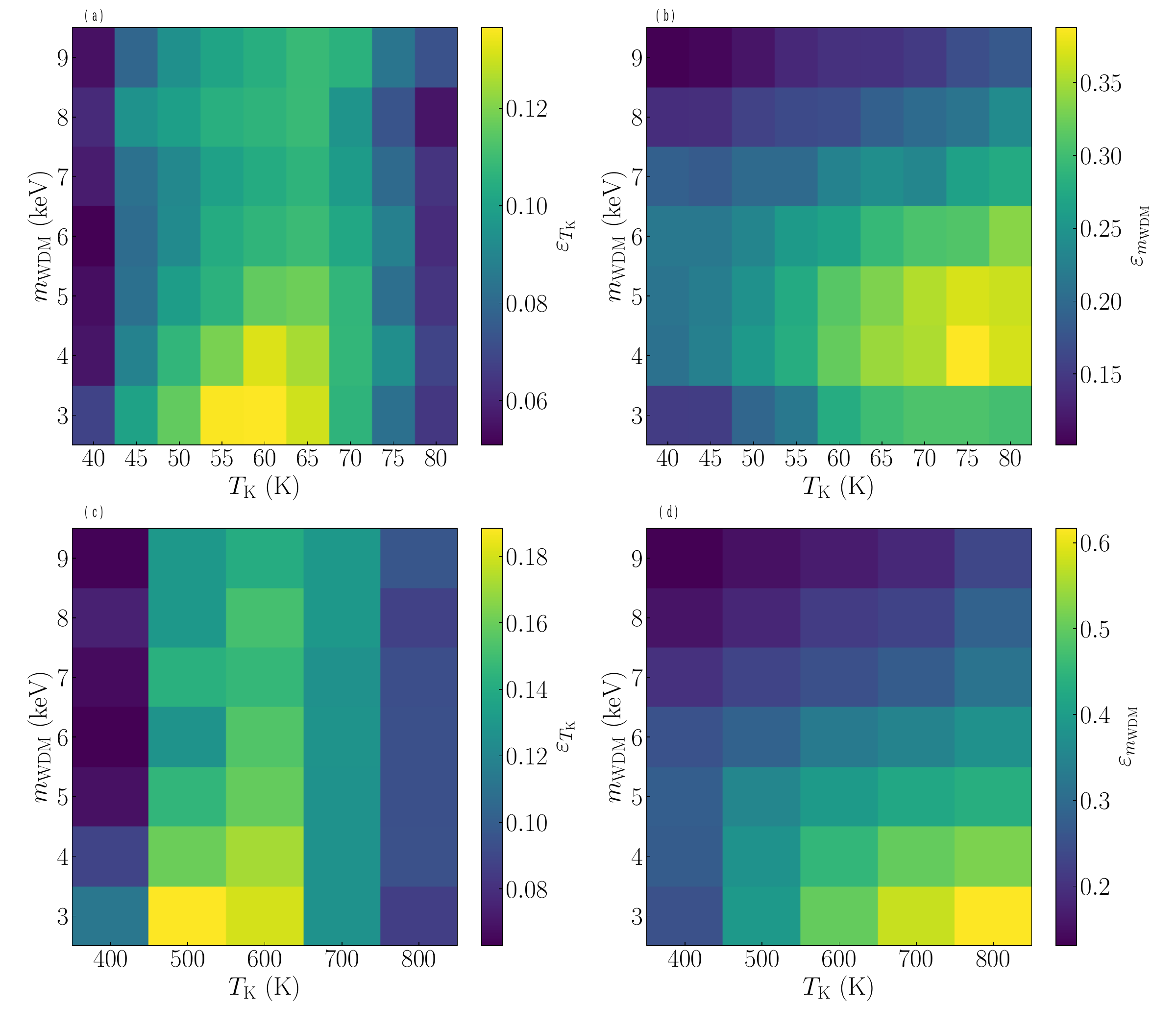}
\caption{\label{sfig6}
  \textbf{The relative error $\varepsilon$ of parameter constraints for the INF for different $T_{\rm K}$ and $m_{\rm WDM}$.}
  \textbf{a, c} Present the temperature constrained precision of the $T_{\rm K}$ using SKA1-LOW and SKA2-LOW, respectively. \textbf{b, d} Present the constrained precision of $m_{\rm WDM}$ using SKA1-LOW and SKA2-LOW, respectively.
  }
\end{sfigure}

\newpage
\begin{sfigure}
\centering
\includegraphics[angle=0, width=16.0cm]{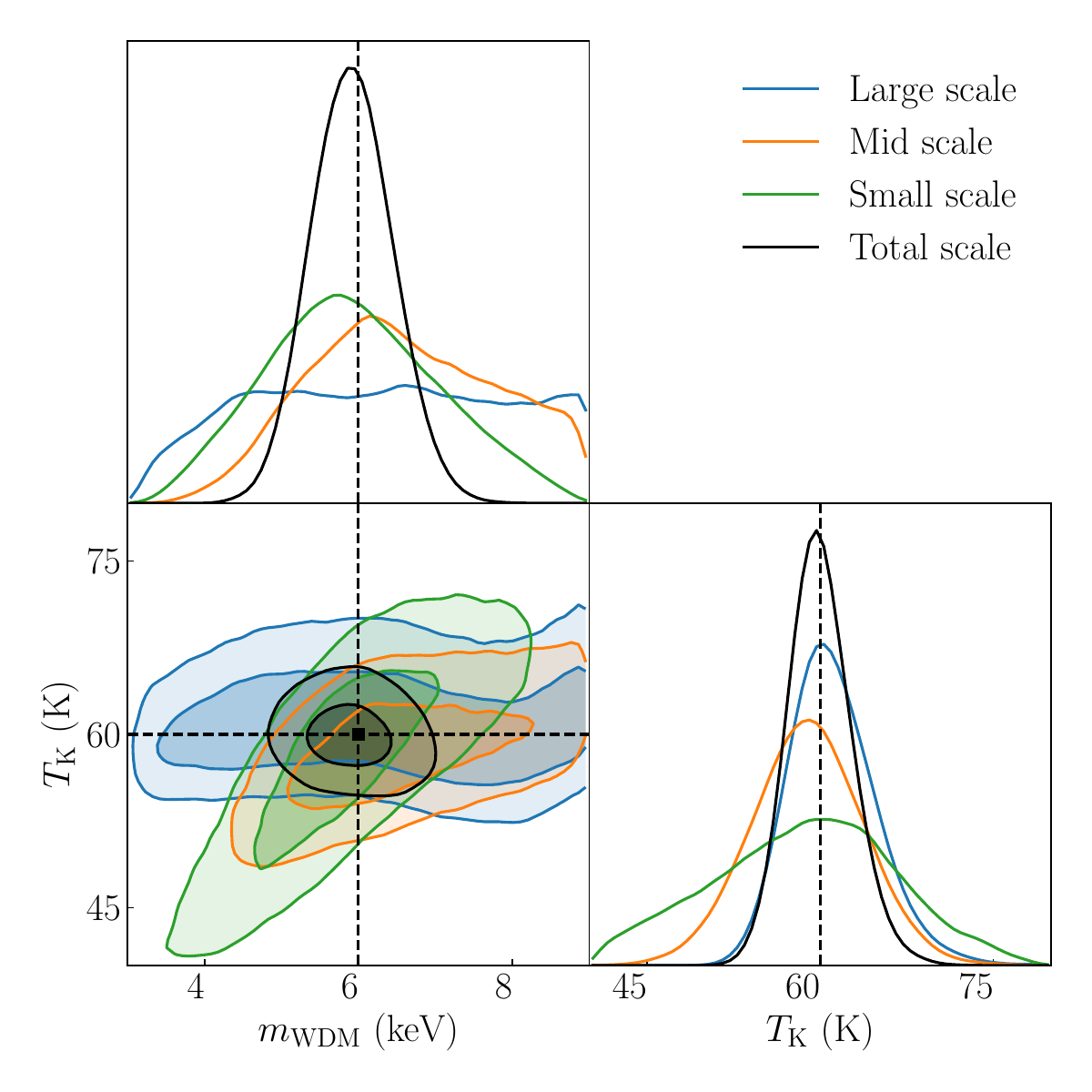}
\caption{\label{sfig7}
  \textbf{The posterior distributions for $m_{\rm WDM}$ and $T_{\rm K}$ using different scales of 1D power spectra with the INF without considering thermal noise.}
  The blue contour is based on the INF for $k$ in [$0.8~\rm Mpc^{-1}$, $3.5~\rm Mpc^{-1}$], the green contour is based on the INF for $k$ in [$5.5~\rm Mpc^{-1}$, $22.0~\rm Mpc^{-1}$], the orange contour is based on the INF for $k$ in [$35.2~\rm Mpc^{-1}$, $144.5~\rm Mpc^{-1}$] and the black contour is based on the INF for total scale. The panel presents the results when $m_{\rm WDM} = 6$ keV and $T_{\rm K} = 60$ K with an integration time of $100$ h using SKA1-LOW.
  }
\end{sfigure}

\newpage
\begin{sfigure}
\centering
\includegraphics[angle=0, width=16.0cm]{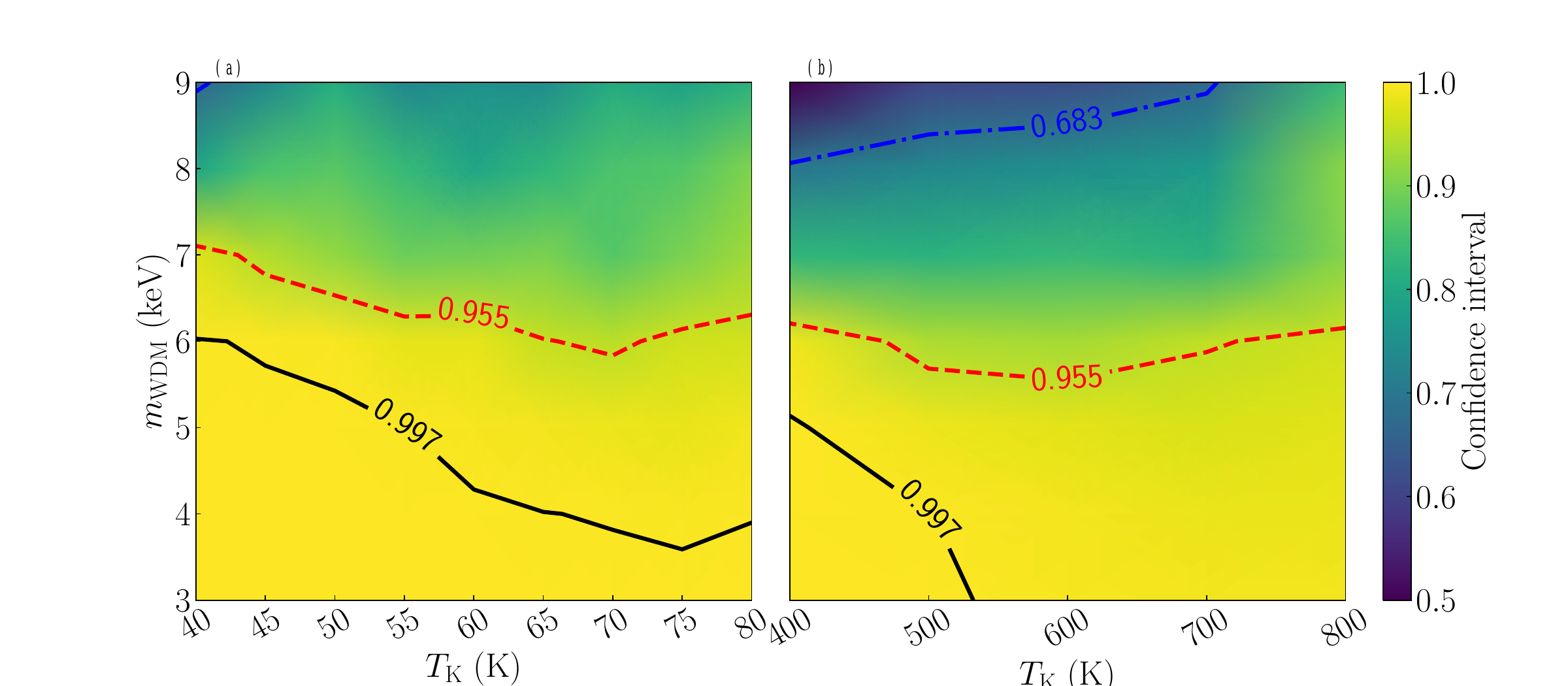}
\caption{\label{sfig8}
  \textbf{Confidence interval distribution for distinguishing WDM from CDM under different $m_{\rm WDM}$ and $T_{\rm K}$.}
  The color represents the confidence intervals, with blue dashed-dotted lines indicating the $1\sigma$ confidence level, dashed red lines indicating the $2\sigma$ confidence level, and solid black lines indicating the $3\sigma$ confidence level. \textbf{a} Presents the scenario with an SKA1-LOW integration time of $100$ h in the case of a weaker heating effect. \textbf{b} Presents the scenario with an SKA2-LOW integration time of $200$ h in the case of a stronger heating effect.
  }
\end{sfigure}

\newpage
\begin{sfigure}
\centering
\includegraphics[angle=0, width=16.0cm]{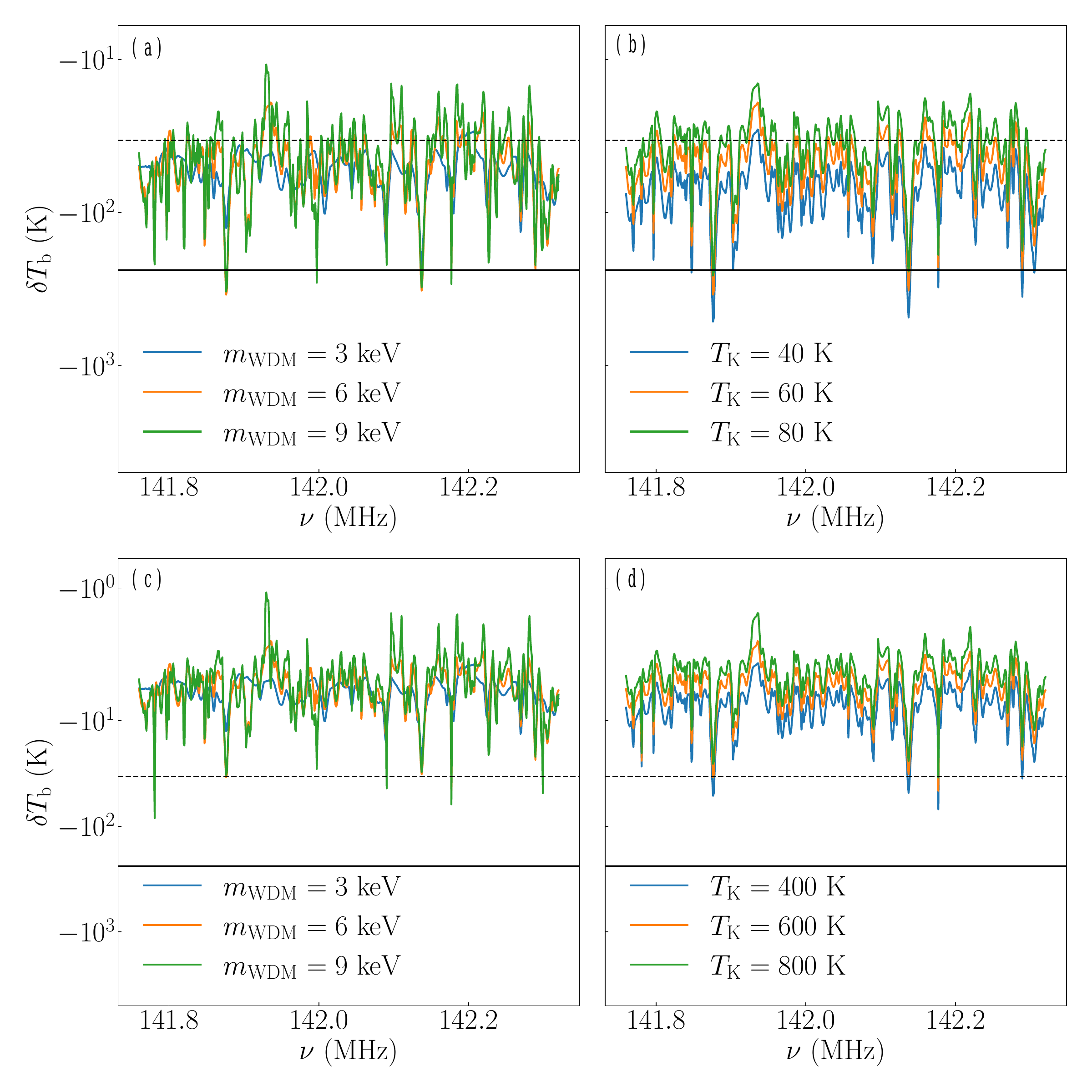}
\caption{\label{sfig9}
  \textbf{Expected brightness temperature of the 21-cm forest.}
 A total of 1 measurements of $10$ comoving megaparsec length segments along a line-of-sight neutral patch against 1 background light source with $S_{150} = 10$ mJy yielded the expected value for the brightness temperature of 21-cm forest at $z = 9$.
 \textbf{a, c} Present the brightness temperature for different $m_{\rm WDM}$ values in the WDM models. The blue, orange, and green curves correspond to $m_{\rm WDM} = 3$ keV, $m_{\rm WDM} = 6$ keV and $m_{\rm WDM} = 9$ keV, respectively. $T_{\rm K}$ in (a) and (c) are 60 K and 600 K, respectively. \textbf{b, d} Present the brightness temperature corresponding to the different $T_{\rm K}$. In all cases, $m_{\rm WDM}$ is set to $6$ keV. The blue, orange, and green curves in \textbf{(b)} correspond to $T_{ \rm K} = 40$ K, $T_{\rm K} = 60$ K and $T_{\rm K} = 80$ K. The blue, orange, and green curves in \textbf{(d)} correspond to $T_{\rm K} = 400$ K, $T_{\rm K} = 600$ K and $T_{\rm K} = 800$ K, respectively. The black solid and dashed lines are thermal noise levels $\delta T^{\mathrm{N}}$ expected for an integration time of $100$ h using SKA1-LOW and an integration time of $200$ h using SKA2-LOW, respectively.
  }
\end{sfigure}


\newpage
\begin{sfigure}
\centering
\includegraphics[angle=0, width=16.0cm]{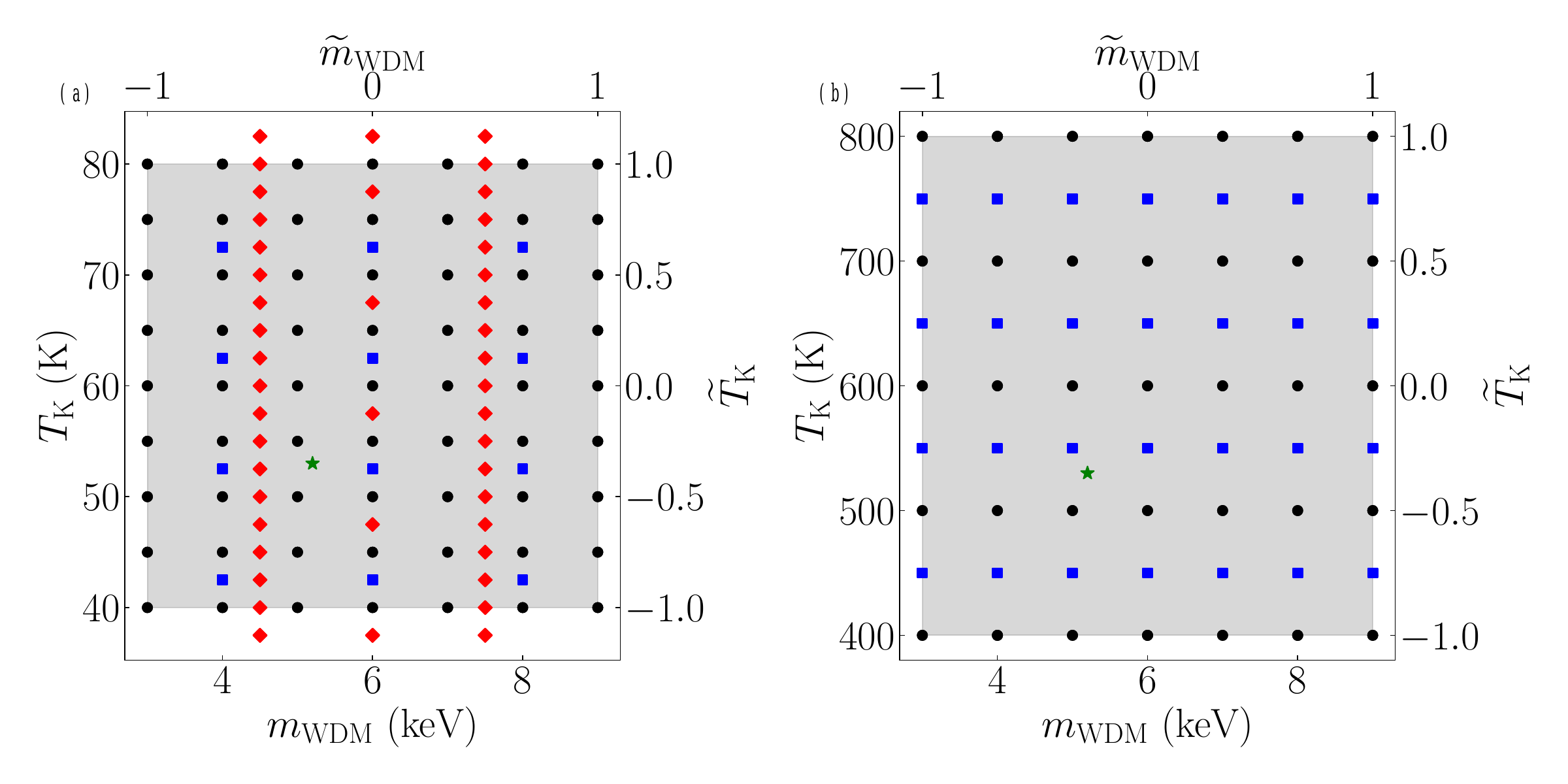}
\caption{\label{sfig10}
  \textbf{The parameter (${m}_{\rm{WDM}}$, ${T}_{\rm{K}}$) distribution of the simulated data.}
  \textbf{a} Presents the distribution of the simulated data with an integration time of $100$ h using SKA1-LOW. \textbf{b} Presents the distribution of the simulated data with an integration time of $200$ h using SKA2-LOW. In both panels, black circle and blue block represent the training set and verification set of the GNF, respectively, while red rhombus corresponds to the testing set of the GNF (only in \textbf{(a)}). Green star represents randomly selected test data points. Both panels display the normalized parameter axes ($\tilde{m}_{\rm{WDM}}$, $\tilde{T}_{\rm{K}}$). For the INF, its testing set corresponds to the untrained portion of the black data set in both panels.
  }
\end{sfigure}

\newpage
\begin{sfigure}
\centering
\includegraphics[angle=0, width=16.0cm]{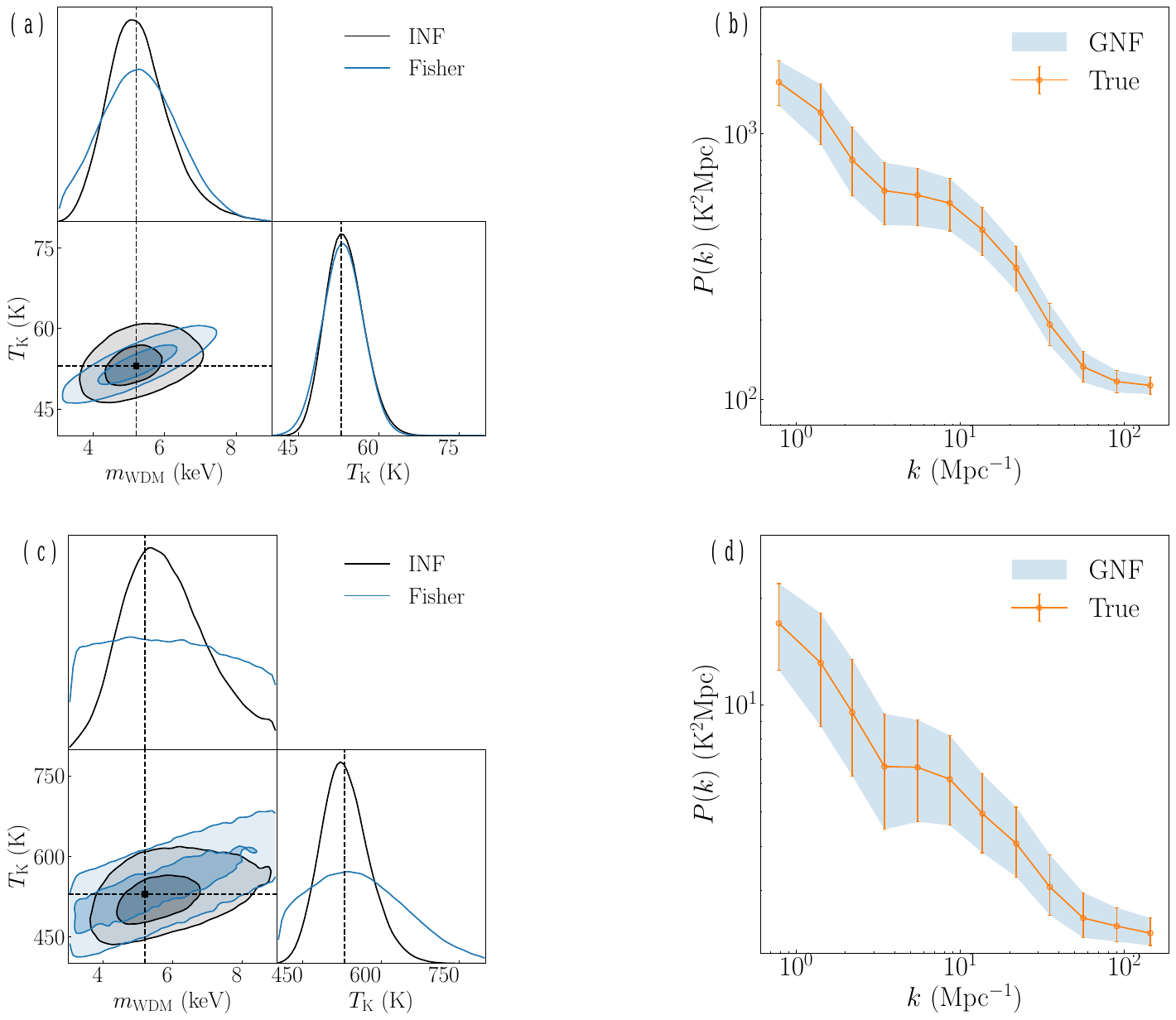}
\caption{\label{sfig11}
\textbf{Same as Figs.~3 and ~6 but testing two random points in the parameter space.}
\textbf{a, b} Present the results based on the fiducial model with$m_{\rm WDM} = 5.2$ keV and $T_{\rm K} = 53$ K with an integration time of $100$ h using SKA1-LOW. \textbf{c, d} Present the results based on the fiducial model $m_{\rm WDM} = 5.2$ keV and $T_{\rm K} = 530$ K with an integration time of $200$ h using SKA2-LOW.
}
\end{sfigure}

\newpage
\begin{sfigure}
\centering
\includegraphics[angle=0, width=16.0cm]{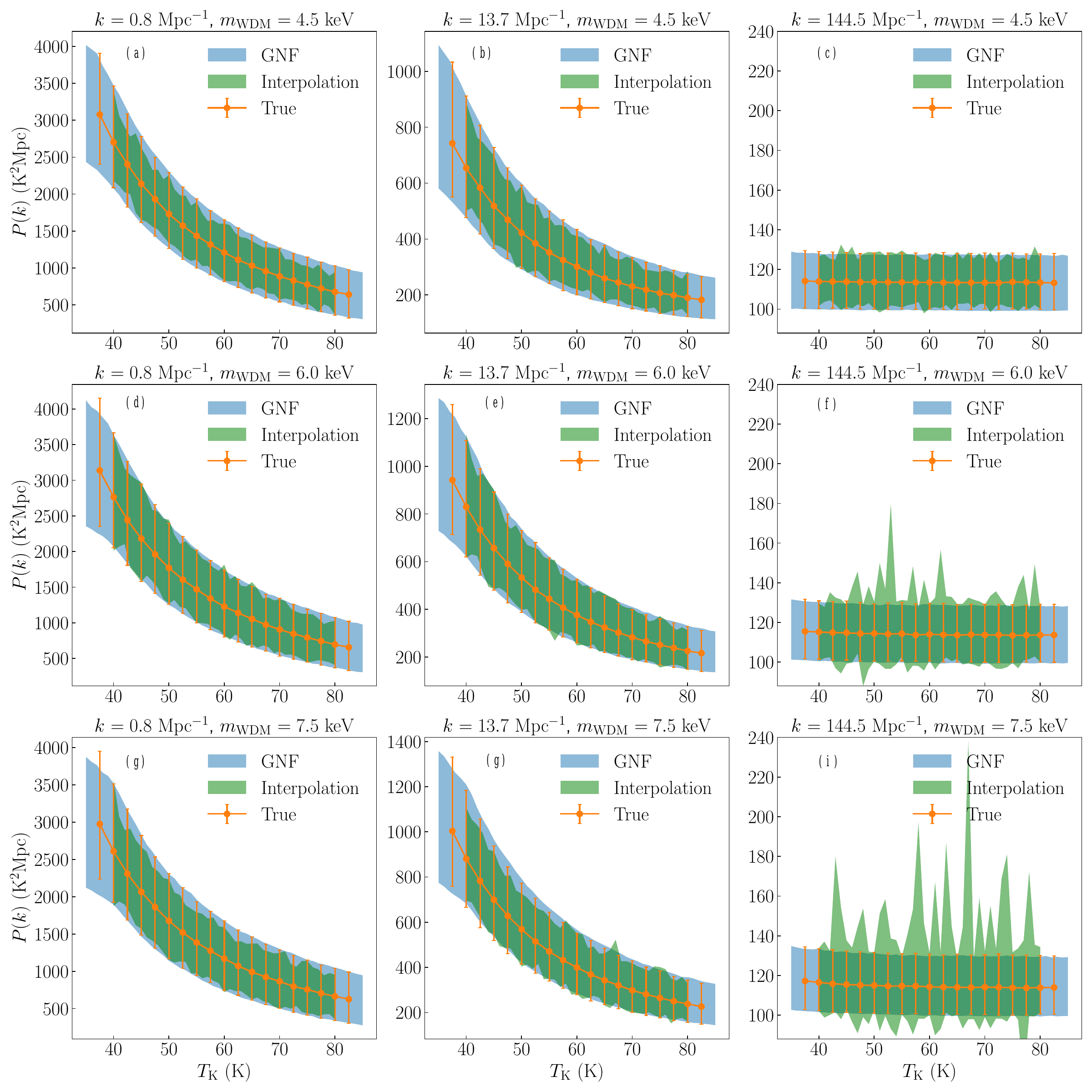}
\caption{\label{sfig12}
 \textbf{The mean values and $1\sigma$ distribution of the 1D power spectra generated by the GNF and linear interpolation methods, compared to simulated 1D power spectra.}
 The data are derived from results of different $T_{\rm K}$ and $m_{\rm WDM}$, with an integration time of $100$ h using SKA1-LOW. \textbf{a - c}, \textbf{d - f}, \textbf{g - i} Present the 1D power spectra of 21-cm forest with $m_{\rm WDM} = 4.5$ keV, $m_{\rm WDM} = 6$ keV and $m_{\rm WDM} = 7.5$ keV, respectively. The 1D power spectrum of 21-cm forest is presented in \textbf{(a, d, g)}, \textbf{(b, e, h)} and \textbf{(c, f, i)} with $k = 0.8~{\rm Mpc}^{-1}$ , $k = 13.7~{\rm Mpc}^{-1}$ and $k = 144.5~{\rm Mpc}^{-1}$, respectively. The blue shaded area represents the $1\sigma$ range of the 1D power spectra generated by the GNF, while the green shaded area represents the $1\sigma$ range of the 1D power spectra generated by linear interpolation. The orange line and bar indicate the mean value and $1\sigma$ error bar of the simulated power spectra.
 }
\end{sfigure}

\begin{stable}
\caption{\label{stable1}
  \textbf{The key hyperparameters of the model algorithm.}
  }
\centering
\setlength\tabcolsep{8pt}
\renewcommand{\arraystretch}{1.5}
\begin{tabular}{c|cccc}
\hline \hline
Hyperparameter      & GNF SKA1 & GNF SKA2& INF SKA1 & INF SKA2 \\
\hline
Learning Rate        & $1\times10^{-4}$     & $1\times10^{-4}$    & $2\times10^{-4}$     & $2\times10^{-4}$     \\
Epochs               & Early Stopping& Early Stopping & $300$      & $250$     \\
Flow Steps           & $6$        & $8$       & $6$        & $6$        \\
Hidden Layer Size    & $256$      & $256$     & $128$      & $256$      \\
Transform Blocks     & $6$        & $8$       & $6$        & $6$         \\
Activation           & ReLU     & ReLU    & ReLU     & ReLU       \\
$L_1$ Regularization & $3\times10^{-6}$     & $3\times10^{-6}$    & $0.0$      & $0.0$         \\
$L_2$ Regularization & $1\times10^{-4}$     & $1\times10^{-4}$    & $0.0$      & $0.0$         \\
Dropout              & $0.0$      & $0.0$     & $0.0$      & $0.0$         \\
Batch Norm           & False    & False   & False    & False        \\
Bins                 & $8$        & $8$       & $8$        &  $8$           \\
Base Transform       &RQ-NSF (AR)&RQ-NSF (AR)&RQ-NSF (AR)&RQ-NSF (AR)\\
\hline \hline
\end{tabular}
\end{stable}

\clearpage
\bibliography{21cmref.bib}

\end{document}